\RequirePackage{fix-cm}
\documentclass{svjour3}                     
\smartqed  

\usepackage{graphicx}
\usepackage{listings}
\usepackage[linesnumbered]{algorithm2e}
\usepackage{subfig}
\usepackage{float}
\usepackage{multirow}
\usepackage{booktabs}
\usepackage{hyperref}
\usepackage{array}
\usepackage{units}
\usepackage{textcomp}
\usepackage{balance}
\usepackage{natbib}
\usepackage{xcolor}
\usepackage[title]{appendix}
\usepackage[normalem]{ulem}
\useunder{\uline}{\ul}{}

\newcommand{\var}{\texttt}

\definecolor{codegreen}{rgb}{0,0.6,0}
\definecolor{codegray}{rgb}{0.5,0.5,0.5}
\definecolor{codepurple}{rgb}{0.58,0,0.82}
\definecolor{backcolour}{rgb}{0.95,0.95,0.92}
\definecolor{bluegreen}{RGB}{51,153,126}	
\definecolor{intenttypecolor}{RGB}{58,162,187}
\definecolor{key-color}{rgb}{0.8, 0.47, 0.196}
\definecolor{xtext-keyword}{RGB}{127, 0, 85}
\definecolor{javadocblue}{rgb}{0.25,0.35,0.75} 
\newcommand{\CNL}{\lstinline[language=OwnExamples, basicstyle=\normalsize\ttfamily,]}
\newcommand{\tinyCNL}{\lstinline[language=OwnExamples, basicstyle=\small\ttfamily,
breaklines=false]}
\newcommand{\exCNL}{\lstinline[language=OwnExamples, basicstyle=\small\ttfamily,
breaklines=true, breakatwhitespace=false]}

\lstdefinelanguage{OwnExamples}{
  keywords={all, each,zero,one,any,none,
     (, ), ,, ,_, ,_and, ,_or, -, ., :, Actors:, After, Before, For, If, Quantifiers:, The, Types:, UI_Component_Types:, Units:, When, Where, While, _and, _or, a, about, accept, accepts, actions, actor, add, adds, after, allow, allows, an, and, and_, any_value, applies, apply, applying, are, as, available, based_on, before, by, calculate, calculates, checking, clicks, compliance, complies, comply, component, conditions, conform, conforms, contain, contains, convention, copies, copy, create, creates, deduct, deducts, default, defined, delete, deletes, described, detect, detects, disable, display, displayed, displays, do, does, double, download, downloads, during, enable, entity, entity_named, equal, equals, every, exclude, excludes, find, finds, finish, finishes, focuses, following, for, format, from, greater, has, have, if, ignore, ignores, in, include, instance, interrupt, into, is, left, less, link, links, migrate, migrates, must, named, not, of, of_type, on, only, opens, or, period, process, properties, property, puts-the-focus, read, reads, receive, receives, referred, reject, rejects, replace, replaces, restore, restores, retrieve, retrieves, right, rule, satisfied, select, selects, send, sends, forward, export, pass, sequence, set, sets, shall, split, splits, standard, start, starting, starts, stop, stops, store, stores, synchronize, synchronizes, 
     than, that, the, then, through, ticks, to, type, unselect, unselects, until, update, updates, upload, uploads, use, uses, using, validate, validates, value, values, when, where, while, with, ANY_OTHER, ID, INT, ML_COMMENT, SL_COMMENT, STRING, WS
  },
   keywordstyle = {\color{xtext-keyword}},
   stringstyle=\color{javadocblue},
   commentstyle=\color{codegreen},
   numberstyle=\tiny\color{codegray},
  comment=[l]{\#},
  morestring=[b]',
  morestring=[b]",
  breaklines=true,
  captionpos=b, 
  basicstyle=\scriptsize\ttfamily,
  prebreak=\raisebox{0ex}[0ex][0ex]{\ensuremath{\hookleftarrow}},
}
\usepackage[colorinlistoftodos]{todonotes}
\usepackage{rotating}
\usepackage{pdflscape}

\newcommand{\ApproachName}{Rimay\xspace}
\newcommand{\ApproachNameNoSpace}{Rimay}

\newcommand*{\affaddr}[1]{#1} 
\newcommand*{\affmark}[1][*]{\textsuperscript{#1}}
\begin{document}

\title{On Systematically Building a Controlled Natural Language for Functional Requirements
}

\titlerunning{On Systematically Building a CNL for Functional Requirements}        

\author{
	Alvaro Veizaga\affmark[1] \and Mauricio Alferez\affmark[1]  \and Damiano Torre\affmark[1] \and
	Mehrdad Sabetzadeh\affmark[2]\,\affmark[1]  \and Lionel Briand \affmark[1]\,\affmark[2]
}

\institute{Alvaro Veizaga \at 
             \email{alvaro.veizaga@uni.lu}           
           \and
           Mauricio Alferez \at
           \email{mauricio.alferez@uni.lu}
           \and
           Damiano Torre \at
           \email{damiano.torre@uni.lu}
           \and
           Mehrdad Sabetzadeh \at
           \email{m.sabetzadeh@uottawa.ca}
           \and
           Lionel Briand \at 
           \email{lionel.briand@uni.lu}\\
           \\
              \affaddr{\affmark[1]~SnT Centre for Security, Reliability and Trust, University of Luxembourg, Luxembourg}\\
              \\
              \affaddr{\affmark[2]~School of Electrical Engineering and Computer Science, University of Ottawa, Canada}\\
}

\date{Received: date / Accepted: date}

\maketitle

\begin{abstract}
\textbf{[Context]}
Natural language (NL) is pervasive in software requirements specifications~(SRSs). However, despite its popularity and widespread use, NL is highly prone to quality issues such as vagueness, ambiguity, and incompleteness. Controlled natural languages (CNLs) have been proposed as a way to prevent quality problems in requirements documents, while maintaining the flexibility to write and communicate requirements in an intuitive and universally understood manner.
\textbf{[Objective]}
In collaboration with an industrial partner from the financial domain, we systematically develop and evaluate a CNL, 
named \ApproachName, intended at helping analysts write functional requirements. \textbf{[Method]}
We rely on Grounded Theory for building \ApproachName and follow well-known guidelines for conducting and reporting industrial case study research.
\textbf{[Results]}
Our main contributions are: (1)~a qualitative methodology to systematically define a CNL for functional requirements; this methodology is general and applicable to information systems beyond the financial domain, (2)~a CNL grammar to represent functional requirements; this grammar is derived from our experience in the financial domain, but should be applicable, possibly with adaptations, to other information-system domains, and (3)~an empirical evaluation of our CNL (Rimay) through an industrial case study.
Our contributions draw on 15 representative SRSs, collectively  containing 3215 NL requirements statements from the financial domain. 
\textbf{[Conclusion]}
Our evaluation shows that \ApproachName is expressive enough to capture, on average, 88\% (405 out of 460) of the NL requirements statements in four previously unseen SRSs from the financial domain.

\end{abstract}

\section{Introduction} \label{sec:Introduction}
Requirements are considered as one of the fundamental pillars of software development. 
For many systems in industry, requirements are predominantly expressed in natural language (NL). Natural language is advantageous in that it can be used in all application domains and understood virtually by all project stakeholders~\citep{PohlKlaus2010Re:f}. Supporting this statement, a study reported that 71.8\% of software requirements specifications (SRSs) are written in NL~\citep{Luisa:2004}. Despite its pervasive use, undisciplined use of NL can bring about a variety of quality issues. Common problems with NL requirements include: poor testability, inappropriate implementation, wordiness, duplication, omission, complexity, vagueness, and ambiguity~\citep{mavin2010big}.

Further, requirements often change throughout a project's lifespan until a consensus is reached among stakeholders. Requirements changes lead to significant additional costs that vary according to the project phase \citep{elizabeth2011requirements}; it has long been known that the cost of fixing problems related to requirements increases rapidly when progressing through the software development phases~\citep{Boehm:2001}. 



The ultimate quality of a software system greatly depends on the quality of its requirements. Empirical evidence shows that the state of  practice for acquiring and documenting requirements is still far from satisfactory~\citep{Sadraei:2007,solemon2009requirements,Young:2015}. Different studies have reported that one of the main causes of software project failures in industry is related to poorly written requirements, i.e., requirements that are unclear, ambiguous, or incomplete~\citep{ahonen2010software,elizabeth2011requirements,ChaosReport}. Poorly written requirements are difficult to communicate and reduce the opportunity to process requirements automatically, for example, to extract models~\citep{arora2015automated} or derive test specifications~\citep{AlferezModels19}.

The problem we address in this article was borne out of a practical need observed across many industrial domains. For example, in the financial domain, the current practice is to write system requirements using a general-purpose text editor without enforcing any requirement structure.
This is the case for our industrial partner, Clearstream Services SA Luxembourg -- a post-trade services provider owned by Deutsche Borse AG. Clearstream reported that several communication problems and delays arise from requirements that are not stated precisely enough, particularly in situations where the project development tasks are divided across several teams in different countries. This problem is compounded by the fact that Clearstream typically has to deal with NL requirements that are written by domain experts (from now on, we refer to them as ``financial analysts''), who do not necessarily possess sufficient expertise in requirements elicitation and definition.

One way to potentially improve the quality of requirements would be to use formal methods. Doing so, however, is not realistic for financial analysts who are unlikely be able to easily express system requirements in a formal notation. In general, these analysts are knowledgeable about the financial domain and related disciplines, e.g., economics, law, management, and accounting, but are, in the majority of the cases, not familiar with discrete mathematics and formal logic. Furthermore, it is not only financial analysts who need to understand the requirements. Other stakeholders at different levels of the organization, e.g., customer service, also need to be able to process the requirements and validate them according to their specific needs~\citep{Dick2017}. 
As a result, there is a tension between the pressure to use NL in practice and the need to be more precise and resorting to formal languages~\citep{yue2011systematic}. Controlled natural languages (CNLs) strike a balance between the usability of NL on the one hand and the rigour of formal methods on the other. A CNL is a set of predefined sentence structures that restrict the syntax of NL and precisely define the semantics of the statements written using these predefined structures~\citep{PohlKlaus2010Re:f}.

In this article, we concern ourselves with developing a CNL for writing requirements for financial applications. We have named our CNL  \ApproachName, which means ``language" in Quechua.
We focus on \emph{functional requirements}, noting that the vast majority of the requirements written by our industrial partner are functional, and that financial analysts find most of the ambiguity and imprecision issues in functional requirements. While our work is grounded in requirements for financial applications, this domain shares many characteristics with other domains where (data-centric) information systems are being developed. As a result, we anticipate that the work presented here, including our methodology, lessons learned, and \ApproachName itself, can contribute to the development of CNLs in other domains.


%

Our investigation is guided by the following research questions (RQs): 
\begin{itemize}
\item\textbf{RQ1: What information content should one account for in the requirements for financial applications?}
In this RQ, we want to identify, in the requirements provided by our industrial partner, the information content used by financial analysts. This information is a prerequisite for the design of the \ApproachName grammar.   
\item\textbf{RQ2: Given the stakeholders, how can we represent the information content of requirements for financial applications?}
After we identify the information content used by our industrial partner to represent requirements, we want to find out the structures of the requirements that our CNL should support. These structures follow recommended syntactic structures and define mandatory and optional information.
\item\textbf{RQ3: How well can \ApproachName express the requirements of previously unseen documents?}
After building our CNL grammar, we need to determine how  well it can capture requirements in unseen SRSs.
\item\textbf{RQ4: How quickly does \ApproachName converge towards a stable state?} We analyze saturation to determine when new SRSs do not entail significant change to \ApproachName, so that it can be considered stable.
\end{itemize}

We use a combination of Grounded Theory and Case Study Research to address the four research questions posed above. The main contributions of this work can be summarized as follows:

\textbf{(1) A qualitative methodology aimed at defining a CNL for functional requirements (RQ1).} 
We rely on Grounded Theory for developing \ApproachName. Our methodology is general and can serve as a good guiding framework for building CNLs systematically.  
We rely on an analysis procedure named \emph{protocol coding}~\citep{saldana2015coding}, which aims at collecting qualitative data according to a pre-established theory, i.e., set of codes. Protocol coding allows additional codes to be defined when the set of pre-established codes is not sufficient. A code in qualitative data analysis is most often a word or short phrase that symbolically assigns a summative, salient, essence-capturing, and/or evocative attribute for a portion of language-based or visual data~\citep{saldana2015coding}. In the context of our article, a code identifies a group of verbs that share the same information content in an NL requirement. As explained in Section \ref{subsub:identify}, most of the codes are pre-existing verb-class identifiers available in a well-known lexicon named VerbNet\footnote{\label{footnote:verbnet}https://verbs.colorado.edu/verbnet/}. In addition, we use WordNet\footnote{https://wordnet.princeton.edu/} to verify the verb senses of the requirements. The fact that we use domain-independent lexical resources and include no keywords specific to the financial domain in \ApproachName, makes our approach more likely to be widely applicable to information systems in general. We conduct our qualitative study on 11 SRSs that contain 2755 requirements in total.

\textbf{(2) A CNL grammar (RQ2) targeting financial applications in particular and information systems in general.} We apply restrictions on vocabulary, grammar, and semantics. 
The \ApproachName grammar accounts for a large variety of system responses and conditions, while following recommended syntactic structures for requirements (e.g., the use of active voice). Also, the \ApproachName grammar defines mandatory information content to enforce the completeness of requirements. In addition to the grammar, we generate a user-friendly and full-featured editor using the language engineering framework Xtext~\footnote{https://www.eclipse.org/Xtext/}. 

\textbf{(3) An empirical evaluation of \ApproachName (RQ3 and RQ4).} We report on a case study conducted within the financial domain. 
We evaluate \ApproachName on four SRSs containing 460 requirements to demonstrate the feasibility and benefits of applying \ApproachName in a realistic context. 
We use \emph{saturation} to find the point in our evaluation where enough SRS content has been analyzed to ensure that \ApproachName is stable for specifying requirements for the financial domain. Furthermore, we use a \emph{z-test for differences in proportions} to confirm that additional enhancements to \ApproachName are unlikely to bring significant benefits.



The article is structured as follows: Section \ref{sec:BackgroundAndRelatedWork} introduces the background and related work. Section~\ref{sec:QualitativeStudy} presents a qualitative study aimed at analyzing the information content in the requirements provided by Clearstream (our industrial partner).
In Section \ref{sec:Grammar}, we describe the details of \ApproachName. Section~\ref{sec:EmpiricalEvaluation} describes a case study that evaluates \ApproachName. Threats to the validity of our results are discussed in Section \ref{sec:ThreatsToValidity}. Section~\ref{sec:PracticalConsiderations} discusses practical considerations and, finally, our conclusions and an outline of future work are provided in Section \ref{sec:Conclusions}.

\section{Background and Related Work} \label{sec:BackgroundAndRelatedWork}
This section reviews the lexical resources we rely on in this work and further discusses related work. 
\subsection{Lexical Resources}
In the next subsections, we discuss WordNet and VerbNet. We use the WordNet dictionary for verb lookup operations and the VerbNet lexicon to cluster verbs with similar semantics into verb classes.

\subsubsection{WordNet}\label{subsec:wordnet} 
WordNet~\citep{miller1995wordnet} is a domain-independent linguistic resource which provides, among several other things, more than 117000 \emph{synsets}. A  \emph{synset}~is a set of synonyms that represent a word sense. Each synset contains information such as  
a definition, an example sentence, and the sense number using the format $word\#sense\ number$. 
For example, in WordNet the verb \emph{create} has six synsets and  synset \#6 is comprised of the following information:  
(a) two synonyms, \emph{produce\#2} and  \emph{make\#6}, (b) the sense definition,  ``create or manufacture a man-made product'', and (c) an example of how to use the verb \emph{create} using synset~\#6, ``We produce more cars than we can sell''. 
In order to develop  \ApproachName, in Section \ref{subsub:identify}, we use WordNet to retrieve the different synonyms and senses of the verbs identified in the NL requirements. 

\subsubsection{VerbNet}\label{subsec:verbnet}
VerbNet~\citep{kipper2000verbnet} is a domain-independent, hierarchical verb lexicon of approximately 5800 English verbs. It clusters verbs into over 270 verb classes, based on their shared syntactic behaviors. 
Each verb in VerbNet is mapped to its corresponding synsets in WordNet,  if the mapping exists. In VerbNet, a verb is always a member of a verb class and each verb class is identified by a unique code composed of a name and a suffix. The suffix reveals the hierarchical level of a verb class, e.g., two of the sub-classes of the root class \emph{multiply-108} are \emph{multiply-108-1} and \emph{multiply-108-2}. In VerbNet, the sub-classes inherit features from the root class and specify further syntactic and semantic commonalities among their verb members. 
For example, each of the sub-classes of \emph{multiply-108} uses the same syntactic structure which is defined as a noun phrase followed by a verb, a noun phrase, and a prepositional phrase. However, each sub-class uses different prepositions in the prepositional phrase. In particular, the subclass \emph{multiply-108-1} has the verb members \emph{divide} and \emph{multiply} and uses the preposition \emph{by} as in the phrase ``I multiplied x by y". The subclass \emph{multiply-108-2} has verb members such as \emph{deduct}, \emph{factor}, and \emph{subtract} and uses the preposition \emph{from} as in the phrase ``I subtracted x from y". 

In Section \ref{subsub:identify}, we describe how we used VerbNet to identify the verb classes of the verbs that we found in our NL requirements.

\subsection{Related Work}
Numerous studies have been conducted with a focus on NL requirements quality improvement.
\citet{PohlKlaus2010Re:f} presents three common techniques for improving the quality of NL requirements by reducing vagueness, incompleteness and ambiguity:
\begin{itemize}
\item[]\textbf{Glossaries}. Requirements glossaries make explicit and provide definitions for the salient terms in a SRS. Requirements glossaries may further provide information about the synonyms, related terms, and example usages of the salient terms~\citep{arora2017automated}. 

\item[]\textbf{Patterns}. They are pre-defined sentence structures that contain optional and mandatory components. Patterns restrict the syntax of the text and are meant to help stakeholders in writing more standardized NL requirements and thus circumventing frequent mistakes. 
\item[]\textbf{Controlled natural languages}. They are considered an extension of the pattern category which, in addition to restricting the syntax (the grammatical structures), also provide language constructs with which it is possible to precisely define the semantics of NL requirements.
\end{itemize}


In this article, we build a CNL to represent functional requirements in the financial domain, but our work likely generalizes to other (data-centric) information systems, noting that \ApproachName does not rely on any domain-specific constructs. 

To identify and synthesize the related work most pertinent to ours, we follow the general principles of conducting systematic literature reviews. Nevertheless, we need to stress that the goal of this article is by no means to report on a fully fledged systematic literature review. Instead, the principles of systematic literature reviews are applied in a lightweight manner as a way to ensure that we have an in-depth understanding of the relevant literature landscape, before we develop and present our own contributions in the next sections.

In our study of related work, we include approaches to describe CNLs and patterns for expressing NL requirements. Note that we only discuss studies using CNLs and patterns, and not studies that focus only on glossaries because they are out of our scope. We considered the approaches that have been published over the past ten years up to September 2019.

The search string we applied is as follows:
\emph{((controlled language OR pattern OR CNL) AND requirement)}.
In order to select the relevant studies, we defined the following inclusion and exclusion criteria. 

\begin{itemize}
\item[]\textbf{Inclusion criteria}
\item Electronic papers focusing on improving writing NL requirements practices through the use of patterns and/or CNLs;
\item Electronic papers written in English;
\item Electronic papers published in peer-reviewed international journals, conferences, and workshops.
\end{itemize}

\begin{itemize}
\item[]\textbf{Exclusion criteria}
\item Electronic papers neither focusing on patterns nor CNLs for improving the writing of NL requirements;
\item Electronic papers that do not provide at least one example of a pattern or a rule for a CNL.
\end{itemize}
We ran our search string over four well-known search engines:  ACM, IEEE, Springer, and ScienceDirect. In order to select a set of primary studies relevant to our topic, we followed a process consisting of three phases. Table~\ref{table:summarySearchResults} summarizes the phases by indicating the number of studies per search engine after each phase. 
\begin{itemize}
\item[]\textbf{P1}: We executed our search string on the four aforementioned search engines and obtained a total of 423 papers.  
\item[]\textbf{P2}: We reduced the initial number of studies by removing duplicates, and by reading the title and in some cases the abstract of the 423 papers obtained after P1. This phase returned a subset of 74 papers.
\item[]\textbf{P3}: We obtained our final set of studies by reading the abstract and the introduction, and by applying the inclusion and exclusion criteria. This resulted in selecting six primary studies from 74 papers (P2) that were analyzed in-depth.  
\end{itemize}

In addition to the six selected primary studies, we consider four other studies: two books~\citep{withall2007software,pohl_rupp_2011}, and two conference papers~\citep{konrad2005real,denger2003higher}, obtained by snowballing the references in the primary studies.
\begin{table}[htpb]
\centering
\caption{Selection of relevant studies}
\label{table:summarySearchResults}
\begin{tabular}{ l | c | c |  c | c | c }
\hline
\textbf{Phase} & \textbf{IEEE} & \textbf{ACM} & \textbf{Springer Link} & \textbf{ScienceDirect} & \textbf{Total} \\
\hline
 P1 &83&70&152&118 &423\\
 P2 &26&8&27&13 &74\\
 P3 &3&1&1&1&6\\ 
\hline
\end{tabular}
\end{table}

Table~\ref{table:relatedWork} outlines the ten studies retained for further analysis. The first column of the table provides a reference to each study. The second column indicates the type of the approach, i.e., Pattern or CNL.  
In order to obtain a more thorough picture of the literature, although our work is focused on functional requirements, our analysis of the related work does not exclude references that exclusively address non-functional requirements. The third column shows the type of the requirements that the approach supports: Functional Requirements (FR), Non-Functional Requirements (NFR), or both.
Additionally, the third column includes the domain in which the patterns and CNLs were created. There are two strands of work: domain-independent and domain-specific (i.e., automotive, business, healthcare, performance, embedded systems, and data-flow reactive systems).

The fourth column indicates whether an empirical study was conducted and evaluated in a systematic manner. The fifth column shows whether the proposed approach was somehow evaluated. 
Finally, the sixth column reports on whether the approach is supported by a tool. 
\begin{table}[htpb]
\centering
\caption{Summary of related work}
\label{table:relatedWork}
\begin{tabular}{ p{2.2cm} | p{1.3cm} | p{2.2cm} | p{1.6cm}  | p{1.4cm} | p{1.1cm}}
\hline
\hfil \textbf{Study } & \hfil \textbf{Type~of}  & \hfil \textbf{Type~of} & \textbf{Systematic}  & \textbf{Evaluation}  & \hfil \textbf{Tool}  \\
\hfil \textbf{Reference}	&\textbf{Approach}&\hfil \textbf{Requirements}	&\hfil \textbf{Study}	 &			& \textbf{Support}\\

\hline
\citet{pohl_rupp_2011} 	&Pattern	&FR (Domain-Independent) &\hfil No	&\hfil No&\hfil No\\  
						&		&&&\\
\citet{mavin2009easy}			&Pattern	&FR (Domain-Independent)	&\hfil No	&\hfil Yes		&\hfil No\\
								&		&&&&\\
\citet{withall2007software}		&Pattern	&Both (Business)	&\hfil No	&\hfil No		&\hfil No\\
								&		&&&&\\
\citet{riaz2014hidden}			&Pattern	&NFR (Healthcare) &\hfil No	&\hfil No		&\hfil Yes\\
								&		&&&&\\
\citet{eckhardt2016challenging}  &Pattern	&NFR (Performance) &\hfil Yes	&\hfil Yes		&\hfil No       \\
								&		&&&&\\
\citet{denger2003higher}  		&Pattern	&FR (Embedded Systems)	&\hfil No	&\hfil Yes		&\hfil No       \\
								&		& &&&\\

\citet{konrad2005real} 			&CNL	&NFR (Automotive)				&\hfil No	&\hfil Yes		&\hfil No       \\ 
 								&		& &&&\\
\citet{post2011applying} 			&CNL	&FR (Automotive) &\hfil No	&\hfil Yes 	&\hfil No         \\ 
								&		&&&&\\
\citet{crapo2017requirements}  	&CNL	&FR (Domain-Independent) 		&\hfil No	&\hfil No 		&\hfil Yes             \\
								&		&		&&&\\
\citet{carvalho2014nat2testscr} &CNL	&Both (Data-Flow Reactive systems) 	&\hfil No	&\hfil No		&\hfil Yes\\

\hline
\end{tabular}
\end{table}

We discuss the selected studies next.
%



\subsubsection{Patterns}
\citet{pohl_rupp_2011} discuss a single pattern to specify functional requirements. The authors claim that the requirements that comply to this pattern are explicit, complete and provide the necessary details to test such requirements. 

\citet{mavin2009easy} define the Easy Approach to Requirements Syntax (EARS), which is a set of five patterns enabling analysts to describe system functions. The authors demonstrate through a case study in the aviation domain that using EARS leads to requirements which are easier to understand and which exhibit fewer quality problems, particularly in relation to ambiguity. 


\citet{withall2007software} identifies 37 patterns to specify structured functional and non-functional requirements for the business domain. The study provides insights regarding the creation and extension of the patterns. 

\citet{riaz2014hidden} define a set of 19 functional security patterns. They provide a tool that assists the user in selecting the appropriate pattern based on the security information identified in the requirements. 

\citet{eckhardt2016challenging} propose patterns to specify performance requirements. The patterns were derived from a content model built from an existing performance classification. \citet{eckhardt2016challenging} define the content elements that a performance requirement must contain to be considered complete. 

\citet{denger2003higher} propose a set of patterns to describe requirements for embedded systems. The patterns were derived from a metamodel that captures several types of embedded-system requirements. The authors validate their patterns through a case study. 

As opposed to the other five studies, only \citet{riaz2014hidden} provide tool support for security patterns to guide analysts in defining requirements. Meanwhile, only \citet{eckhardt2016challenging} follow a systematic process to develop a framework for the creation of   
performance requirements patterns, and presented a well-defined evaluation of their approach. 

\subsubsection{Controlled Natural Languages}

\citet{konrad2005real} provide a restricted natural language for the automotive and appliance domains, enabling analysts to express precise qualitative and real-time properties of  systems. 
They evaluate their approach through a case study. 

\citet{post2011applying} identify three new rules that extend the approach proposed by~\citet{konrad2005real} to express requirements in the automotive domain. They also validated their rules through a case study. 

\citet{crapo2017requirements} propose the Semantic Application Design Requirements Language which is a controlled natural language in English for writing functional requirements. Their language supports the mapping to first-order logic. 
\citet{carvalho2014nat2testscr} propose a CNL called SysReq-CNL that allows analysts to describe data-flow requirements. Their sentence rules are nonetheless not mapped onto any formal semantics. None of the above approaches have been empirically evaluated. 

To summarize, no previous strand of work describes a systematic process to build CNL grammar rules.  Further, only~\citet{crapo2017requirements} and~\citet{carvalho2014nat2testscr} provide tool support to assist analysts with specifying requirements. 

\subsubsection{Differences Between the Related Work and Our Approach}
No other work, in our knowledge, follows a systematic process for creating and evaluating a CNL to specify functional requirements, either in the financial domain (the main focus of our investigation) or any other domain. More precisely, our work differs from the existing work in the following respects: (a) we derive \ApproachName from the analysis of a large and significant number of requirements from the financial domain; (b) we create \ApproachName by following a rigorous and systematic process; (c) we evaluate \ApproachName through a case study based on industrial data while following  empirical guidelines for conducting Case Study Research~\citep{Runeson:2012}; and (d) we fully operationalize \ApproachName through a usable prototype tool.
\section{Qualitative Study}\label{sec:QualitativeStudy}
In this section, we report on a qualitative study aimed at characterizing the information content found in the functional NL requirements provided by Clearstream. In the following, every time we speak of ``requirements'', we mean functional NL requirements. 
First, we describe the context of the qualitative study along with the criteria used to select SRSs. 
Then, we present the analysis procedure of our qualitative study where we show the codes that identify different groups of requirements. Each group of requirements is characterized by different information content. In this work, information content refers to the meaning assigned to the text of the requirements. The result of the analysis procedure is a grammar that defines the syntax of a CNL that is able to specify all the information content found in the analyzed requirements.

\subsection{Research Question}
The goal of this qualitative study is to answer the following research question: \emph{\textbf{RQ1: What information content should one account for in the requirements for financial applications?}}
RQ1 aims to identify the mandatory and optional information content used by Clearstream to describe requirements. 
This is essential in order to design a CNL that will help financial analysts write requirements that are as complete and as unambiguous as possible. 

\subsection{Study Context and Data Selection}

\label{subsec:StudyContexDataSelection}
\sloppypar
We conducted this study in collaboration with Clearstream Services SA Luxembourg, which is a securities services company with 2500 customers in 110 countries. We validated our results and conclusions with a team of experts at the company. The team was composed of eight financial analysts: (a) two were \emph{senior} financial analysts that had more than 20 years of experience in specifying requirements for the financial domain; (b) four were \emph{mid-career} financial analysts with more than 10 (but less than 20) years of expertise in the financial domain, and (c) two were \emph{junior} financial analysts with between 2 to 5 years of experience in the financial domain. This validation activity was performed in an iterative and incremental manner with face-to-face plenary bi-weekly sessions with the team of experts, with each of these sessions lasting between 2 to 3 hours. 
 
Clearstream is continuously delivering new software projects in the financial domain and employs English as the primary language for specifying requirements.

Among all those available in Clearstream, we selected SRSs which: 
(a)~belong to recently concluded projects, 
(b)~contain at least 15 requirements,
(c)~contain requirements written  in English, and
(d)~are written by different financial analysts.
The senior financial analysts from Clearstream selected 11 representative SRSs according to the four criteria defined above. Each one of the SRSs contained the following types of information: business context, goals and objectives, project scope, current and future overview, general information (e.g., glossary, related documentation, acronyms and abbreviations), and Unified Modeling Language (UML) diagrams for the high-level functional decomposition of the systems 
and requirements. 
In total, the 11 SRSs contained 2755 requirements. 

\subsection{Analysis Procedure}
\begin{figure}[htpb]
\centering
 \centerline{\includegraphics[width=0.8\linewidth]{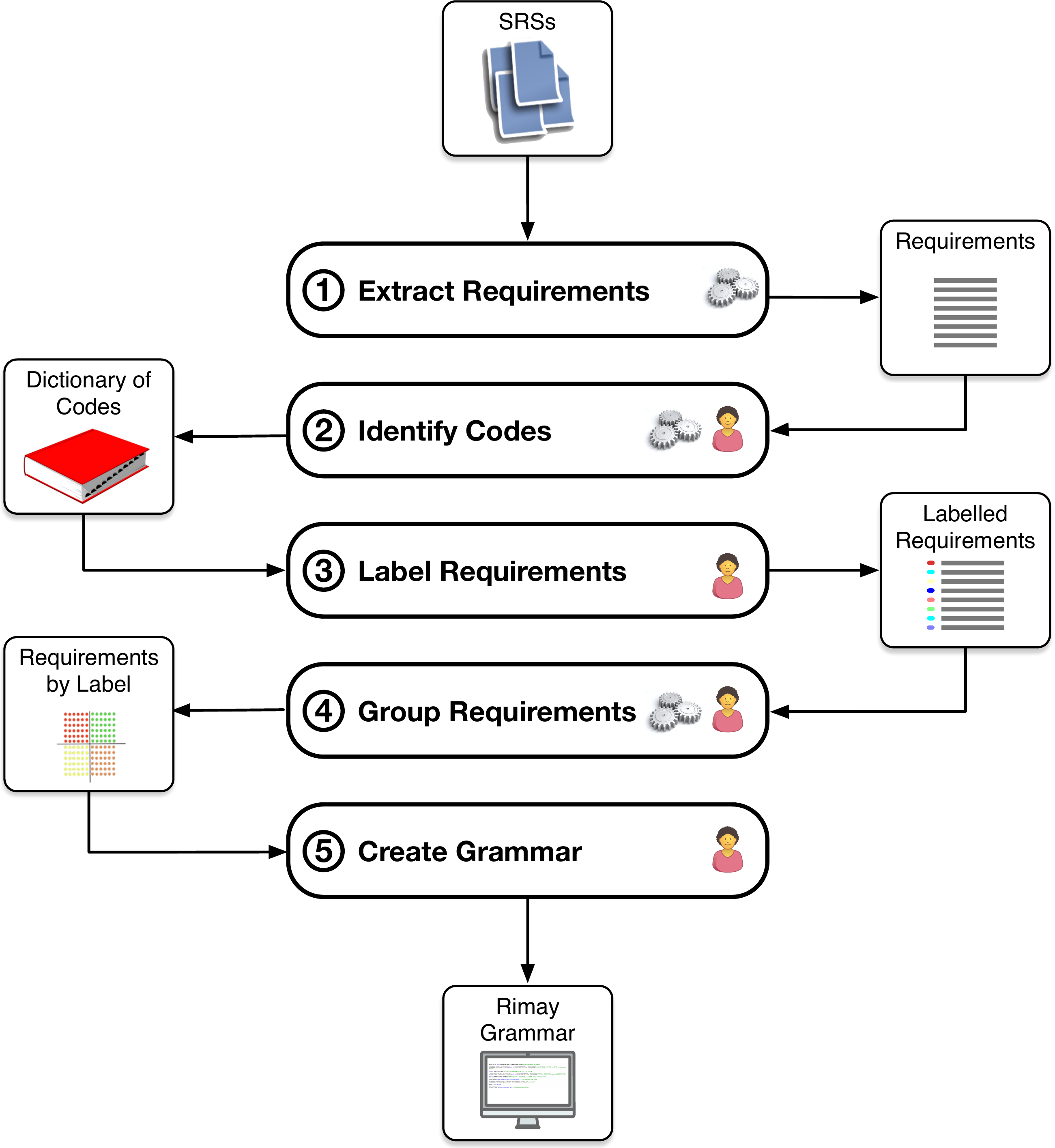}}
\caption{Overview of our analysis procedure}
\label{fig:CreationProcess}
\end{figure}

Figure~\ref{fig:CreationProcess} shows an overview of our analysis procedure. In Step~1, we first extracted 2755 requirements from 11 SRSs. 
In Step~2, we identified a dictionary of 41 codes from the extracted requirements. 
For example, the code \emph{send\_11.1} identifies five verbs used in the extracted requirements: ``return", ``send", ``forward", ``pass", ``export" and ``import"(Table~\ref{table:CodingResultsVNetQualitative} and Table~\ref{table:CodingResultsProposedQualitative} shows the 41 codes and verbs identified in our qualitative study and the evaluation). 
Our analysis procedure for identifying the codes followed \emph{protocol coding}~\citep{saldana2015coding}, which is a method for collecting qualitative data according to a pre-established theory, i.e., a set of codes. Protocol coding allows additional codes to be defined when the set of pre-established codes are not sufficient. 
In Step~3, two annotators (first and second authors of this article) labeled the extracted requirements with one or more of the codes discovered in the previous step.
In Step~4, we grouped the extracted requirements by their labels. The purpose of grouping requirements is to ease the identification of common information content to create grammar rules. For example, all the requirements that use the verbs members of the code \emph{send\_11.1} share the semantic roles \texttt{INITIAL LOCATION} (a place where an event begins or a state becomes true) and \texttt{DESTINATION} (a place that is the end point of an action and exists independently of the event).
In Step~5, we iteratively created and integrated the grammar rules into \ApproachName. 
Each of the five steps in Figure~\ref{fig:CreationProcess} shows one or two icons denoting whether a given step was carried out (1) automatically (i.e., the three gears icon), (2) manually (i.e., the human icon), or (3) semi-automatically (i.e., both icons). 
The next subsections describe in details Steps~1~to~5.

\subsubsection{Extract Requirements (Step~1)}
We read the 11 SRSs and extracted 2755 requirements. In our case, all the requirements were written in tables in which all the requirements were clearly identified and distinguished from other information.

Table \ref{table:ExtractRequirements} shows three requirements extracted from a SRS. The column ``Id'' identifies the requirements, the column ``Description" contains the original text of the requirements, and the column ``Rationale" presents the reasoning behind the creation of a given requirement. 

\begin{table}[htpb]
\centering
\caption{Three requirements extracted from a SRS during Step~1 of Figure~\ref{fig:CreationProcess}}
\label{table:ExtractRequirements}
\begin{tabular}{p{2.1cm} | p{4.4 cm} | p{4.4cm} } 
\hline
\hfil \textbf{Id} & \hfil \textbf{Description} & \hfil \textbf{Rationale} \\
\hline

TNG.INPUT.010   &  If the message contains ``FISN", then the System must ignore the message. &  FISN is an official ISO Standard created to enhance the quality of financial messaging.      \\
TRAN.0030&The System must regenerate the outbound XML according to the new XML specification ``SR2017". & The previously created orders, which their status are activated, must be changed to comply with the new XML specification.\\
 Data.SAA.060& The data of the System older than 13 months must be archived for at least 10 years.          & This requirement complies to a legal rule.
\\
\hline

\end{tabular}
\end{table}

\subsubsection{Identify Codes (Step~2)}\label{subsub:identify}
The requirements specify the expected system behavior using verb phrases, e.g., ``\emph{send} a message" and ``\emph{create} an instruction". 
We used the verb lexicon named VerbNet~(Section \ref{subsec:verbnet}) to identify the codes from our SRSs. 
Subsection \ref{subsec:createGrammar} will explain in details how, by using verb classes, we obtain the grammar rules of \ApproachName. 



We followed a semi-automated process to identify codes and their corresponding verbs. From the 41 codes that we proposed in this qualitative study, 32 codes (78\%) correspond to verb class ids from VerbNet (referred to thereafter as \emph{VerbNet codes}), and nine (22\%) are codes that we proposed because they were missing from VerbNet but were needed to analyze the requirements.
We use below the following terms to describe this process: 
\begin{itemize}
\item \texttt{REQS:} Set of requirements to analyze.
\item \texttt{LEMMAS:} List of lemmas found in the action phrases of \texttt{REQS}.
\item \texttt{CODES:} Dictionary of codes and their corresponding verb members found during our analysis procedure. There are two types of codes: VerbNet codes and codes proposed by us. 
\item \texttt{AUX:} Auxiliary list of the lemmas that are not members of any code in \texttt{CODES}.
\item \texttt{SYNS:} Dictionary of lemmas and their corresponding applicable synonyms.
\item  \texttt{VN:} Read-only dictionary of all the publicly available VerbNet codes and their corresponding verb members.
\end{itemize}
\begin{figure}[htpb]
\centering 
\centerline{\includegraphics[width=0.87\linewidth]{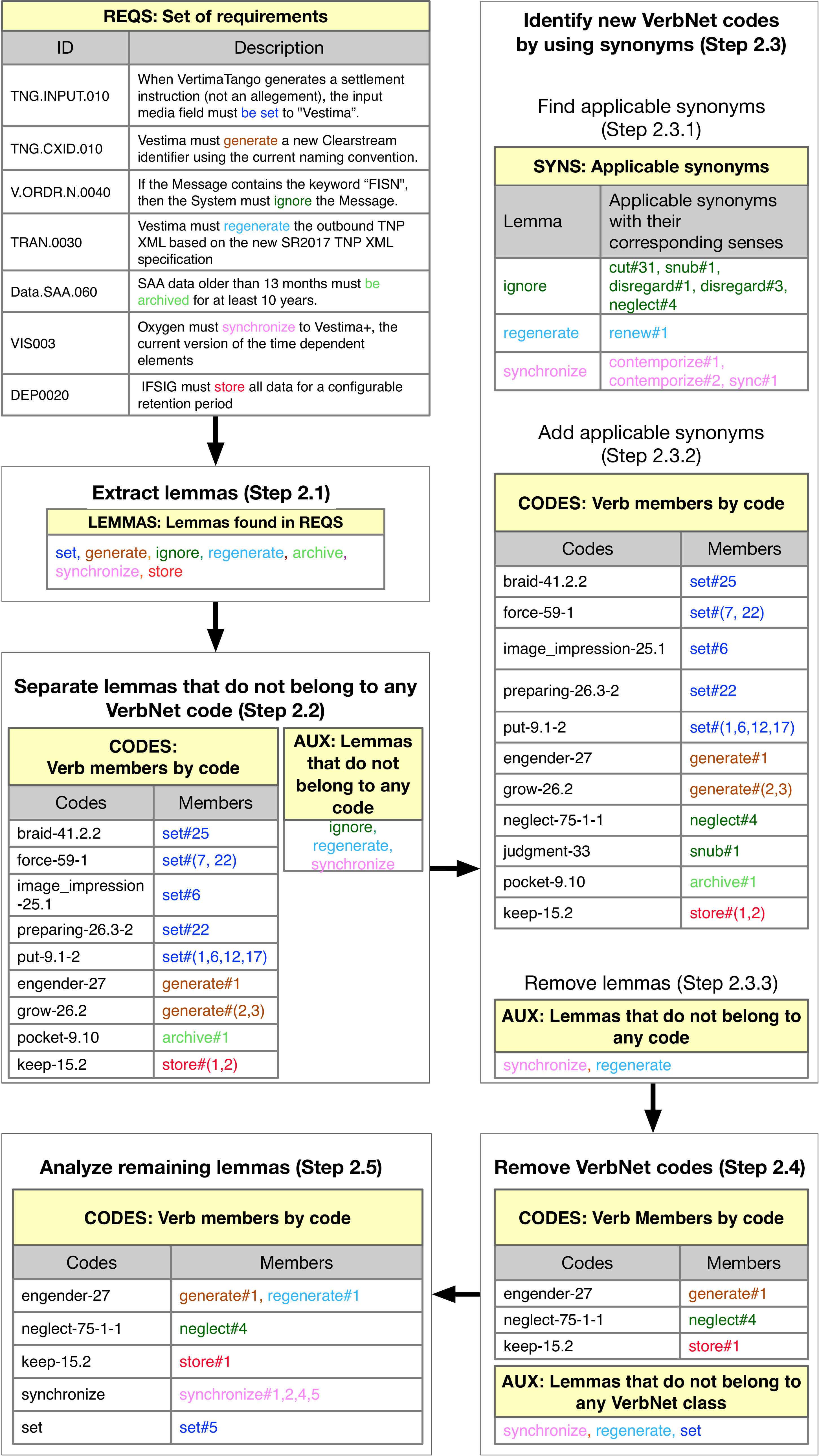}}
\caption{Identify codes (Step~2)}
\label{fig:Step2IdentifyVerbClasses}
\end{figure}

In Figure~\ref{fig:Step2IdentifyVerbClasses}, we show a running example of our process to identify the codes. The process steps are as follows: 
\paragraph{Extract lemmas (Step 2.1).} 
We extracted the verbs of each requirement in \texttt{REQS} (upper-left corner of Figure~\ref{fig:Step2IdentifyVerbClasses}) to obtain lemmas. A lemma is the base form of the verb. For example, from ``archived", the lemma is ``archive". We stored the resulting lemmas in \texttt{LEMMAS}. 

\paragraph{Separate lemmas that do not belong to any VerbNet code (Step 2.2).} We retrieved for every lemma in \texttt{LEMMAS} its corresponding VerbNet codes from \texttt{VN}. We stored these VerbNet codes and their corresponding lemmas (including their sense number, depicted as a number after the symbol \emph{\#}) in \texttt{CODES}.
For example, the key-value pair \emph{\{engender-27, generate\#1\}} in \texttt{CODES} of Figure~\ref{fig:Step2IdentifyVerbClasses}~(Step~2.2) means that the lemma \emph{generate} (Step~2.1 of Figure~\ref{fig:Step2IdentifyVerbClasses}) with the sense number one (i.e., ``bring into existence") is a member of the VerbNet code \emph{engender-27}. 

If a lemma in \texttt{LEMMAS} was not a member of any VerbNet code in \texttt{VN}, we added it to an auxiliary list of lemmas named \texttt{AUX}. For example, in  Figure~\ref{fig:Step2IdentifyVerbClasses}~(Step~2.2) we added to \texttt{AUX} the lemmas \emph{ignore}, \emph{regenerate} and \emph{synchronize} that were not identified in \texttt{VN}, but were found in the analyzed requirements.

\paragraph{Identify new VerbNet codes by using synonyms (Step 2.3).} We analyzed the synonyms and senses of the lemmas in \texttt{AUX} to discover new VerbNet codes that can be added to  \texttt{CODES}. 
We describe this process in more details as follows:

\paragraph{Find applicable synonyms (Step 2.3.1).} We used WordNet to retrieve all the synonyms of each auxiliary lemma in~\texttt{AUX}. We stored in  \texttt{SYNS} only the synonyms whose senses match the sense of an auxiliary lemma as used in ~\texttt{REQS}. 

As an example, Table~\ref{table:Senses} shows the list of synonyms of the lemma \emph{regenerate}, which is one of the lemmas in \texttt{AUX} shown in  Figure~\ref{fig:Step2IdentifyVerbClasses}~(Step~2.2).
The synonyms in Table~\ref{table:Senses} are grouped according to the sense numbers of the lemma \emph{regenerate}, namely 1, 3, 4 and 9 (according to WordNet, the verb \emph{regenerate} has nine senses, but Table~\ref{table:Senses} only shows the senses that have at least one synonym). From the four senses in Table~\ref{table:Senses}, we chose the ones that match the sense of the verb \emph{regenerate} used in \texttt{REQS}. In this case, we chose sense number~1 since it was the only sense that was applicable to the requirements. 
Finally, we store in \texttt{SYNS} the synonyms and their chosen sense numbers. In the case of the lemma \emph{regenerate}, we only added \emph{renew\#1} to \texttt{SYNS}.


\begin{table}[htpb]
\caption{Senses and synonyms of the verb \emph{regenerate} retrieved from WordNet.}
\label{table:Senses}
\begin{tabular}{p{1.2cm} | p{4.8 cm} | p{3.3cm} | p{0.9cm} } 
\hline
\hfil \textbf{Sense} &\hfil \textbf{Sense Definition} & \hfil \textbf{Synonyms and} & \textbf{Chosen  }\\
\textbf{Number}& &\hfil \textbf{Their Sense Number} & \textbf{Sense?}\\

\hline
\hfil 1 & Reestablish on a new, usually improved, basis or make new or like new & renew\#1 & \hfil Yes \\

\hfil 3 &Bring, lead, or force to abandon a wrong or evil course of life, conduct, and adopt a right one & reform\#2, reclaim\#3, rectify\#3 & \hfil No\\
\hfil 4 & Return to life, get or give new life or energy & restore\#2, rejuvenate\#4 & \hfil No \\
\hfil 9 & Restore strength & revitalize\#1 & \hfil No\\
\hline
\end{tabular}
\end{table}

\paragraph{Add applicable synonyms (Step 2.3.2).}
We retrieved, for every synonym in \texttt{SYNS}, its corresponding VerbNet codes from \texttt{VN}. Then, we stored the retrieved VerbNet codes and the corresponding synonym (including the sense number) in \texttt{CODES}. For example, given that the synonym \emph{neglect} (Step~2.3.1 of Figure~\ref{fig:Step2IdentifyVerbClasses}) with sense number four (i.e., \emph{neglect\#4}) is a member of the VerbNet code \emph{neglect-75-1-1}, we created the key-value pair \emph{\{neglect-75-1-1, neglect\#4\}} in \texttt{CODES}~(Step~2.3.2 of Figure~\ref{fig:Step2IdentifyVerbClasses}).
If none of the synonyms of a lemma is a member of any code in \texttt{VN}, then we move the lemma from \texttt{SYNS} to \texttt{AUX}.
For example, if the synonym is \emph{renew\#1} and it is not a member of any VerbNet code in \texttt{VN}, if it is a synonym of \emph{regenerate} we then move \emph{regenerate} from \texttt{SYNS} to \texttt{AUX}. 

\paragraph{Remove VerbNet codes (Step 2.4).}\label{subsec:removeCodes}
In this step, our goal is to remove the VerbNet codes (from \texttt{CODES}) that are either not relevant to the SRSs in the financial domain or redundant. 
We performed this step during several offline validation sessions. Each session was attended by three to four financial analysts with the presence of at least one senior and one mid-career financial analyst.

At the end of Step~2.4 (Figure~\ref{fig:Step2IdentifyVerbClasses}), we went from 11 to three VerbNet codes (i.e., a reduction of 72,7\%).
Considering all the VerbNet codes used during this qualitative study, not only the 11 VerbNet codes shown in Step~2.4 in Figure~\ref{fig:Step2IdentifyVerbClasses}, we decreased the number of VerbNet codes from 158 to 32 (i.e., a reduction of 79,7\%).
The two strategies that we employed to reduce VerbNet codes are as follows:

\begin{itemize}
\item \emph{Strategy 1. Discard redundant verbs.} 
For example, between the verbs \emph{archive} and \emph{store}, we discard the verb \emph{archive} because the verb \emph{store} is more frequent and both verbs are semantically similar.
\item \emph{Strategy 2. Discard verbs that do not have applicable senses.} 
For example, the VerbNet code \emph{image\_impression-25.1} (Step~2.3.2 of Figure~\ref{fig:Step2IdentifyVerbClasses}) involves only the member \emph{set\#6} whose sense is defined by WordNet as: ``a relatively permanent inclination to react in a particular way". Since this latter sense is not used in \texttt{REQS}, we finally discarded \emph{image\_impression-25.1} from \texttt{CODES}. After applying this strategy, if a verb was discarded from \texttt{CODES}, we added only its lemma to \texttt{AUX} for further manual analysis as we explain next in Step 2.5. For example, given that the verb \emph{set} was discarded from \texttt{CODES}, we added its lemma (e.g., only the word \emph{set} without sense\emph{\#}) to \texttt{AUX}.
\end{itemize}

\paragraph {Analyze remaining lemmas (Step 2.5).} In this step, we manually checked in WordNet if the senses of the remaining lemmas in \texttt{AUX} could be included in \texttt{CODES}. This step was carried out with the help of two senior and two mid-career financial analysts from Clearstream. 
We updated \texttt{CODES} when we identified an appropriate sense in WordNet that referred to one of the remaining lemmas. For example, in Figure~\ref{fig:Step2IdentifyVerbClasses}, we created the code \emph{set} with a member \emph{set\#5} whose sense is used in \texttt{REQS}, and updated the VerbNet code \emph{engender-27} with the member \emph{regenerate\#1}.

\paragraph{Coding results.}\label{subsec:codingResults}
Tables~\ref{table:CodingResultsVNetQualitative} and \ref{table:CodingResultsProposedQualitative} present the resulting codes identified during our qualitative study described in Section~\ref{subsub:identify} (``Identify Codes" (Step~2)). We finally obtained 41~codes, where 32 were obtained from VerbNet and nine were proposed by us.

Table~\ref{table:CodingResultsVNetQualitative} provides the 32 VerbNet codes and their members. The first column of the table lists the codes, where each code is composed of a class name and a hierarchy level. The second column shows the verb members related to the code.  

Similarly, Table~\ref{table:CodingResultsProposedQualitative} shows the nine codes that we proposed. The first column of the table lists the codes and the second column provides the verb members associated to the code. 

\begin{table}[htpb]
\centering
\caption{VerbNet codes identified during our qualitative study}
\label{table:CodingResultsVNetQualitative}

\begin{tabular}{p{2.9cm} | l | p{6.1cm} }
\hline

\multicolumn{2}{c|}{\textbf{Codes}} 	 & \multicolumn{1}{c}{\multirow{2}{*}{\hfil \textbf{Members} }}  			\\[0.1cm] \cline{1-2}
\multicolumn{1}{c|}{\hfil \textbf{Class name}} & \multicolumn{1}{c|}{\hfil \textbf{Hierarchy}} & \multicolumn{1}{c}{}     \\ 
\multicolumn{1}{c|}{} & \multicolumn{1}{c|}{\hfil \textbf{Level }} & \multicolumn{1}{c}{}\\

\hline 
admit & 65 & exclude\tabularnewline
advise & 37.9-1 & instruct\tabularnewline
allow & 64.1 & allow, authorize \tabularnewline
beg & 58.2 & request\tabularnewline
begin & 55.1-1 & begin\tabularnewline
concealment & 16-1 & hide\tabularnewline
contribute & 13.2 & restore\tabularnewline
create & 26.4 & compute, publish\tabularnewline
enforce & 63 & enforce \tabularnewline
engender & 27 & create, generate\tabularnewline
exchange & 13.6 & replace\tabularnewline
forbid & 67 & prevent\tabularnewline
herd & 47.5.2 & aggregate\tabularnewline
involve & 107 & include\tabularnewline
keep & 15.2 & store\tabularnewline
limit & 76 & limit, restrict, reduce\tabularnewline
mix & 22.1-2 & add\tabularnewline
mix & 22.1-2-1 & link\tabularnewline
neglect & 75-1-1 & neglect, ignore\tabularnewline
obtain & 13.5.2 & accept, receive, retrieve\tabularnewline
other\_cos & 45.4 & close\tabularnewline
put & 9.1 & insert\tabularnewline
reflexive appearance & 48.1.2 & display, show\tabularnewline
remove & 10.1 & extract, remove, delete \tabularnewline
say & 37.7-1 & report, propose\tabularnewline
see & 30.1-1 & detect \tabularnewline
send & 11.1 & return, send, forward, pass\tabularnewline
shake & 22.3-2-1 & concatenate \tabularnewline
throw & 17.1 & discard\tabularnewline
transcribe & 25.4 & copy\tabularnewline
turn & 26.6.1 & convert, change, transform\tabularnewline
use & 105 & apply \tabularnewline
\hline 
\multicolumn{3}{c}{Total: 32} \tabularnewline
\hline
\end{tabular}
\end{table}

\begin{table}[htpb]
\centering
\caption{Codes proposed during the qualitative study}
\label{table:CodingResultsProposedQualitative}
\begin{tabular}{ p{2.5cm} |  p{3.0cm}  }
\hline
\textbf{Codes} & \textbf{Members} \\  [0.1cm]
\hline
cancel 	&cancel          	 \\
enable disable &enable, disable\\
get from	&download	\\
interrupt 	&interrupt		\\
migrate 	&migrate		\\
select unselect  &select, unselect 	\\
synchronize	&synchronize		\\
update &update		\\
validate  		&validate, check 	\\
\hline
\multicolumn{2}{c}{Total:  9}\tabularnewline
\hline
\end{tabular}
\end{table}

\subsubsection{Label Requirements (Step~3)}
In Step~3  (Figure~\ref{fig:CreationProcess}), two annotators (the two first authors of this article) manually labeled the requirements extracted in Step~1 with one or more of the codes identified in Step~2. The labeling process required to (a) read the requirements and identify the verbs used in the system response of the requirements, (b) attempt to match the identified verbs with members of the codes found in Step~2, and (c) when there is a match, label the requirement with the corresponding code. 

We describe below the three activities of the labeling process for requirement DEP0020 in \texttt{REQS} shown in Figure~\ref{fig:Step2IdentifyVerbClasses}:``IFSIG must store all data for a configurable retention period". Specifically,  (a) we identified that the verb used in the system response is \emph{store}, (b) we detected that \emph{store} matches one of the members of the VerbNet code \emph{keep-15.2}, and (c) we labeled the requirement with the VerbNet code \emph{keep-15.2}.



\subsubsection{Group Requirements (Step~4)}
In Step~4 (Figure~\ref{fig:CreationProcess}), we grouped and copied the labeled requirements to different spreadsheets based on their labels. The purpose of having the requirements grouped by label is to make it easier for us to identify common information content among them.

\subsubsection{Create Grammar (Step~5)}\label{subsec:createGrammar}
\begin{figure}[htpb]
\centering  
\centerline{\includegraphics[width=0.87\linewidth]{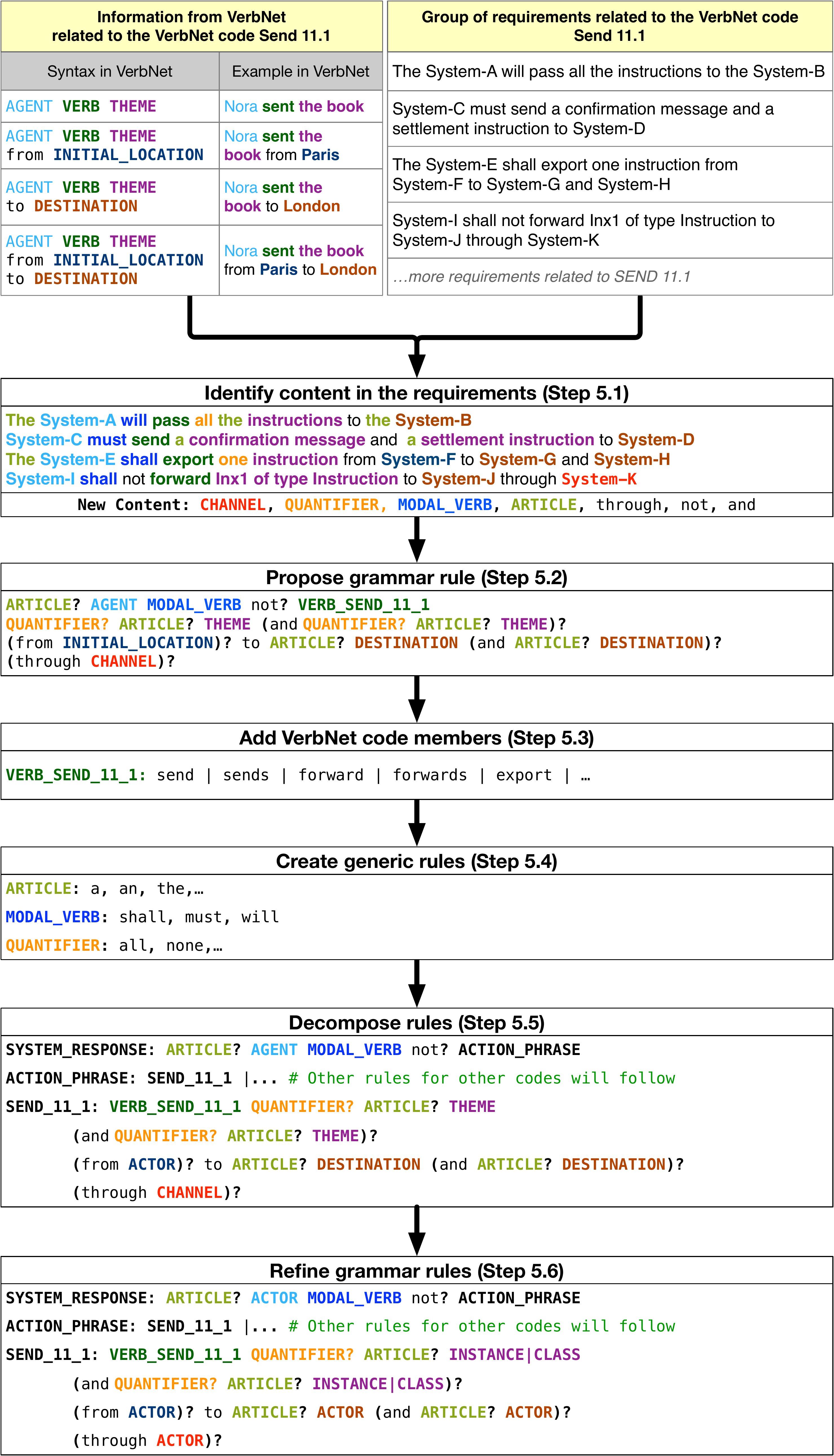}}
\caption{Obtaining CNL grammar rules from requirements related to the VerbNet code \emph{Send 11.1}}
\label{fig:ClassToRule}
\end{figure}

In Step~5 (Figure~\ref{fig:CreationProcess}) we created the grammar of \ApproachName to capture relevant information content from the requirements. 
Figure~\ref{fig:ClassToRule} shows the steps that we carried out to create grammar rules for the VerbNet code \emph{Send 11.1} (Table~\ref{table:CodingResultsVNetQualitative}). The box in the upper-right corner of Figure~\ref{fig:ClassToRule} shows four examples of requirements related to the VerbNet code \mbox{\emph{Send 11.1}} that will be used to illustrate this step. The same sub-steps (i.e., from 5.1 to 5.6) were carried out for the rest of the codes presented in Table~\ref{table:CodingResultsVNetQualitative} and Table~\ref{table:CodingResultsProposedQualitative}. 

\paragraph{Identify content in the requirements (Step 5.1).} In this step we identify semantic roles and keywords in the requirements. VerbNet provides the syntax and the examples that show most of the semantic roles and the keywords (e.g., the prepositions) related to the VerbNet codes in Table~\ref{table:CodingResultsVNetQualitative}.
For example, the box in the upper-left corner of Figure~\ref{fig:ClassToRule} shows the syntax and examples related to the VerbNet code \mbox{\emph{Send 11.1}}. The syntax contains the prepositions \texttt{from} and \texttt{to}, and the semantic roles \texttt{AGENT} (a participant that initiates an action), \texttt{THEME} (an entity which is moved by an action, or whose location is described), \texttt{INITIAL\_LOCATION} (a place where an event begins or a state becomes true) and \texttt{DESTINATION} (a place that is the end point of an action and exists independently of the event). 

In Figure~\ref{fig:ClassToRule}, we use different colors to show the correspondence between the semantic roles and the parts of the requirements that represent the semantic roles. When some content in the requirements was not related to any VerbNet semantic role, we proposed a new semantic role to identify that content. For example, in Step 5.1 of Figure~\ref{fig:ClassToRule}, we proposed the new semantic role \texttt{CHANNEL} to identify the content in the phrase \mbox{``through System-K"}.


\paragraph{Propose grammar rule (Step 5.2).}
Based on the syntax provided by VerbNet, we defined the order of appearance of the content, and its repetition in \ApproachName. The symbols \texttt{?}, \texttt{*} and \texttt{+} indicate that the users of \ApproachName can repeat what is before the symbol at most once, any number of times, and at least once, respectively. Step 5.2 in Figure~\ref{fig:ClassToRule} shows that the grammar rule for the VerbNet code \emph{Send 11.1} contains keywords such as (i) connectors (\emph{and} and \emph{or}), (ii) prepositions shown in the VerbNet syntax (\emph{from} and \emph{to}), (iii) prepositions related to new content (\emph{through}) and (iv) the negation of a modal verb (\emph{not}). 

\paragraph{Add VerbNet code members (Step 5.3).} 
We added a complete list of all the members of each VerbNet code related to its corresponding rule. For example, \emph{forward} and \emph{send} are two of the members of the VerbNet code \emph{Send~11.1} that we added to its corresponding rule \texttt{VERB\_SEND\_11\_1}. We also added the conjugated forms of the verbs to the rule (e.g., \emph{forwards}, \emph{sends}).
	
\paragraph{Create generic rules (Step 5.4).} 
We created the rules related to the generic English grammar, e.g., we created the rules \texttt{ARTICLE}, \texttt{MODAL\_VERB}, and \texttt{QUANTIFIER}.

\paragraph{Decompose rules (Step 5.5).} We decomposed the grammar rules created in Step~5.2 to make them easier to understand and reuse. For example, we decomposed the example rule in Step 5.2 into three rules: \texttt{SYSTEM\_RESPONSE}, \texttt{ACTION\_PHRASE}, and \texttt{SEND\_11\_1}. 
\paragraph{Refine grammar rules (Step 5.6).}With the help of four financial analysts (including one senior and one mid-career financial analyst), 
we replaced some of the semantic role names with other ones that were more familiar to both financial analysts and engineers. In our case, financial analysts and engineers working for Clearstream were familiar with the UML~\citep{UML25}. For example, in the grammar rules \texttt{SYSTEM\_RESPONSE} and \texttt{SEND\_11\_1} (Step 5.4 in Figure~\ref{fig:ClassToRule}), we chose to replace the role \texttt{AGENT} with \texttt{ACTOR}, because an agent can be represented as an UML actor, i.e., a role played by a human user or a system who initiates and carries out an event or action.

\paragraph{\textbf{Method.}}
The method that we used to create \ApproachName was iterative and incremental. 
This means that we first followed Steps~5.1~to~5.6 in Figure~\ref{fig:ClassToRule} to create the grammar rules related to one of the groups of requirements produced in Step~4 of Figure~\ref{fig:CreationProcess}. 
Second, we generated a requirements editor using Xtext.
Third, we used the generated editor to rephrase the requirements in the first requirements group to test the grammar and its corresponding editor. We tested that our grammar and the editor were expressive enough to allow us to write all the information content for the first group of requirements. If the grammar was not expressive enough, we analyzed and extended the grammar, regenerated the editor and verified the requirements until there were no errors in all the rephrased requirements. For each remaining requirements groups produced in Step~4 (Figure~\ref{fig:CreationProcess}), we repeated Steps~5.1~to~5.6 as performed for the first requirements group.


\noindent\fbox{%
\parbox{\textwidth}{\textbf{Answer to RQ1:} Following a systematic and repeatable process, we  identified 41 codes, which in our context, are groups of verbs that convey the same information in NL requirements. We created grammar rules for all the codes identified, thus covering all the information content found in a large and representative set of functional requirements in financial applications. We anticipate that our approach, being general in nature, should be applicable to other domains as well. 
    }%
}
\section{Controlled Natural Language for Functional Requirements}\label{sec:Grammar}

In this section, we describe how a requirement is structured in \ApproachName in order to answer \emph{\textbf{RQ2: ``Given the stakeholders, how can we represent the information content of requirements for financial applications?"}}.  

The rule \texttt{REQUIREMENT} shown in Listing~\ref{GeneralRules} provides the overall syntax for a requirement in \ApproachName. The rule shows that the presence of the \texttt{SCOPE} and \texttt{CONDITION\_STRUCTURES} is optional, but the presence of an \texttt{ACTOR}, \texttt{MODAL\_VERB} and a \texttt{SYSTEM\_RESPONSE} is mandatory in all requirements.

\begin{lstlisting}[frame=single,
label={GeneralRules},
language=OwnExamples,
linerange=1-2,
caption={Overall syntax of \ApproachName}
%\color{red}More general grammar rules of our CNL}
]
REQUIREMENT: SCOPE? CONDITION_STRUCTURES? ARTICLE? ACTOR MODAL_VERB not?  SYSTEM_RESPONSE.
CONDITION_STRUCTURES: CONDITION_STRUCTURE (,? (and|or) CONDITION_STRUCTURE)*, then?
\end{lstlisting}

In a requirement, an actor is expected to achieve a system response if some conditions are true. 
An actor is a role played by an entity that interacts with the system by exchanging signals, data or information~\citep{UML25}. 
Moreover, requirements written in \ApproachName may have a scope to delimit the effects of the system response. 
One example of a requirement in \ApproachName is: ``\CNL{For all the depositories, System-A must create a MT530 transaction processing command}". The requirement has a scope (\CNL{For all the depositories}), does not have any conditions, and has an actor (\CNL{System-A}) and a system response (\CNL{create a MT530 transaction processing command}). 

Throughout this section, we simplify the description of \ApproachName by considering that the keywords are not case-sensitive. Also, we use grammar rules that are common in English such as \texttt{MODAL\_VERB} (e.g., shall, must) and \texttt{MODIFIER} that includes articles (e.g., a, an, the) and quantifiers (e.g., all,  none,  only one, any). 
Subsections~\ref{subsec:conditionStructures} and \ref{subsec:systemResponse} will explain the \texttt{CONDITION\_STRUCTURES} and \texttt{SYSTEM\_RESPONSE}, respectively.

\subsection{Condition Structures}\label{subsec:conditionStructures}
The grammar rule named \texttt{CONDITION\_STRUCTURE} shown in Listing~\ref{ConditionStructures} defines different ways to use system states, triggering events, and features, to express conditions that must hold for the system responses to be triggered.

\noindent\begin{minipage}{\linewidth}
\begin{lstlisting}[frame=single,
label={ConditionStructures},
language=OwnExamples,
linerange=4-10, 
caption={Condition structures}]
CONDITION_STRUCTURE: WHILE_STRUCTURE|WHEN_STRUCTURE|WHERE_STRUCTURE|IF_STRUCTURE|TEMPORAL_STRUCTURE
WHILE_STRUCTURE: While PRECONDITION_STRUCTURE
WHEN_STRUCTURE: When TRIGGER
WHERE_STRUCTURE: Where TEXT #TEXT is a feature expression
IF_STRUCTURE: If PRECONDITION_STRUCTURE|TRIGGER
TEMPORAL_STRUCTURE: (Before|After)(TIME|TRIGGER)
\end{lstlisting}
\end{minipage}

The condition structures \texttt{WHILE}, \texttt{WHEN}, \texttt{WHERE} and \texttt{IF} that we use in our grammar are inspired by the EARS template~\citep{mavin2009easy}. EARS is considered by practitioners as beneficial due to the low training overhead and the quality and readability of the resultant requirements~\citep{DBLP:conf/re/MavinWGU16}. Additionally, we proposed the rule \texttt{TEMPORAL\_STRUCTURE} to be used when the system responses are triggered before or after an event. 
Below, we describe the types of \texttt{CONDITION\_STRUCTURE} used in \ApproachName:
\begin{itemize}
	\item The \texttt{WHILE\_STRUCTURE} is used for system responses that are triggered while the system is in one or more specific states.
	\item The \texttt{WHEN\_STRUCTURE} is used when a specific triggering event is detected at the system boundary.
	\item The \texttt{WHERE\_STRUCTURE} is used for system responses that are triggered only when a system includes particular features. The features are described in free form using the rule \texttt{TEXT}.
	\item The \texttt{IF\_STRUCTURE} is used when a specific triggering event happens or a system state should be hold at the system boundary before triggering any system responses. 
\end{itemize}

The rule \texttt{CONDITION\_STRUCTURE} shown in Listing~\ref{ConditionStructures} allows combining condition structures using logical operators. We can, for example, combine the \texttt{IF} and \texttt{WHEN} structures using the operator \CNL{and} in the structure ``\CNL{If PRECONDITION_STRUCTURE and when TRIGGER}" to separate the conditions in which the requirement can be invoked (i.e., the preconditions) and the event that initiates the requirement (i.e., the trigger). 

Figure~\ref{fig:ExampleConditionStructure} depicts examples of the \texttt{WHEN\_STRUCTURE}, \texttt{TEMPORAL\_STRUCTURE}, and \texttt{IF\_\_STRUCTURE}. 

Listing~\ref{Trigger} shows the grammar rules \texttt{TRIGGER} and \texttt{PRECONDITION\_STRUCTURE} referenced by the condition structures in Figure~\ref{fig:ExampleConditionStructure}. 
\begin{lstlisting}[frame=single,
label={Trigger},
language=OwnExamples,
linerange=1-7,
caption={Trigger and precondition structure}]
TRIGGER: MODIFIER? ACTOR ACTIONS_EXPRESSION
ACTION: ((do|does) not )? ACTION_PHRASE
PRECONDITION_STRUCTURE: ITEMIZED_CONDITIONS | CONDITIONS_EXPRESSION 
TIME: TEXT UNIT #TEXT describes an amount of time units
ITEMIZED_CONDITIONS: the following conditions are satisfied:
	HYPHEN CONDITION((,|,and|and) 
	HYPHEN CONDITION)+
\end{lstlisting}

The rule \texttt{TRIGGER} in Listing~\ref{Trigger} defines that a triggering event is always caused by an \texttt{ACTOR} that performs some actions. The actions performed by the actor are defined by the rule \texttt{ACTIONS\_EXPRESSION} which enables the combination of any number of actions using logic connectors to express complex system events. The \texttt{WHEN\_STRUCTURE} in Figure~\ref{fig:ExampleConditionStructure} shows an example of a trigger composed of an actor and an action expression: \mbox{``\CNL{System-B receives an email alert from System-A}"}. 


The rule \texttt{PRECONDITION\_STRUCTURE} in Listing~\ref{Trigger} gives freedom for the users to decide how to describe conditions. The rule \texttt{ITEMIZED\_CONDITIONS} (Listing~\ref{Trigger}) is appropriate for writing long lists of conditions that must evaluate to True. Conversely, the rule \texttt{CONDITIONS\_EXPRESSION} (Listing~\ref{Trigger}) is suitable for only one condition, multiple conditions combined with logical operators, or parentheses that denote priority in the evaluation order of operations. The \texttt{IF\_STRUCTURE} in Figure~\ref{fig:ExampleConditionStructure} shows examples of non-itemized and itemized conditions.

\subsection{Conditions}
In the previous subsection, we introduced the rule \texttt{PRECONDITION\_STRUCTURE} to specify conditions.
This rule is composed of operands and operators which are described as follows.

\subsubsection{Operands}\label{subsubsec:Operands}
The operands are represented by the rules \texttt{ACTOR}, \texttt{CLASS}, \texttt{PROPERTY}, \texttt{INSTANCE}, \texttt{ELEMENT} and \texttt{TEXT}. The meaning of the operands is the same as in the UML~\citep{UML25}, therefore an \emph{Actor} specifies a role played by the user or another system that interacts with our system. The \emph{Class} represents a domain concept (e.g., Instruction). A \emph{Property} represents the attributes of the \emph{Class}. An \emph{Instance} represents a specific realization of a \emph{Class} and an \emph{Element} is a constituent of a model. 

The users of \ApproachName can use the dot notation to refer to a property of a class, e.g.,``\CNL{Instruction.Settlement_Date}". 
In the cases where there is only one instance of a class in a requirement, the users do not need to declare any instance. For example, given that in Figure~\ref{fig:ExampleConditionStructure} there is only one instance of an instruction, we used ``\CNL{Instruction}" instead of ``\CNL{Inx1 of type Instruction}".

\subsubsection{Operators} 
\ApproachName uses the following families of operators and its negative forms: 
\begin{itemize}
	\item \texttt{COMPARE}, such as ``\CNL{equals to}", ``\CNL{less or equal to}", etc.,
	\item \texttt{CONTAINS} such as ``\CNL{has}", ``\CNL{contains}", etc., 
	\item \texttt{OTHER OPERATORS} such as ``\CNL{is available}" 
\end{itemize}
An example of a condition that  conforms to \ApproachName is: ``\CNL{Inx1 of type Settlement_Instruction has Status and Status is equal to Valid}". This condition uses operators of type \texttt{CONTAINS} and \texttt{COMPARE}.
\begin{landscape}\centering
\vspace*{\fill}
\begin{figure}[htpb]
	{\includegraphics[scale=0.6]{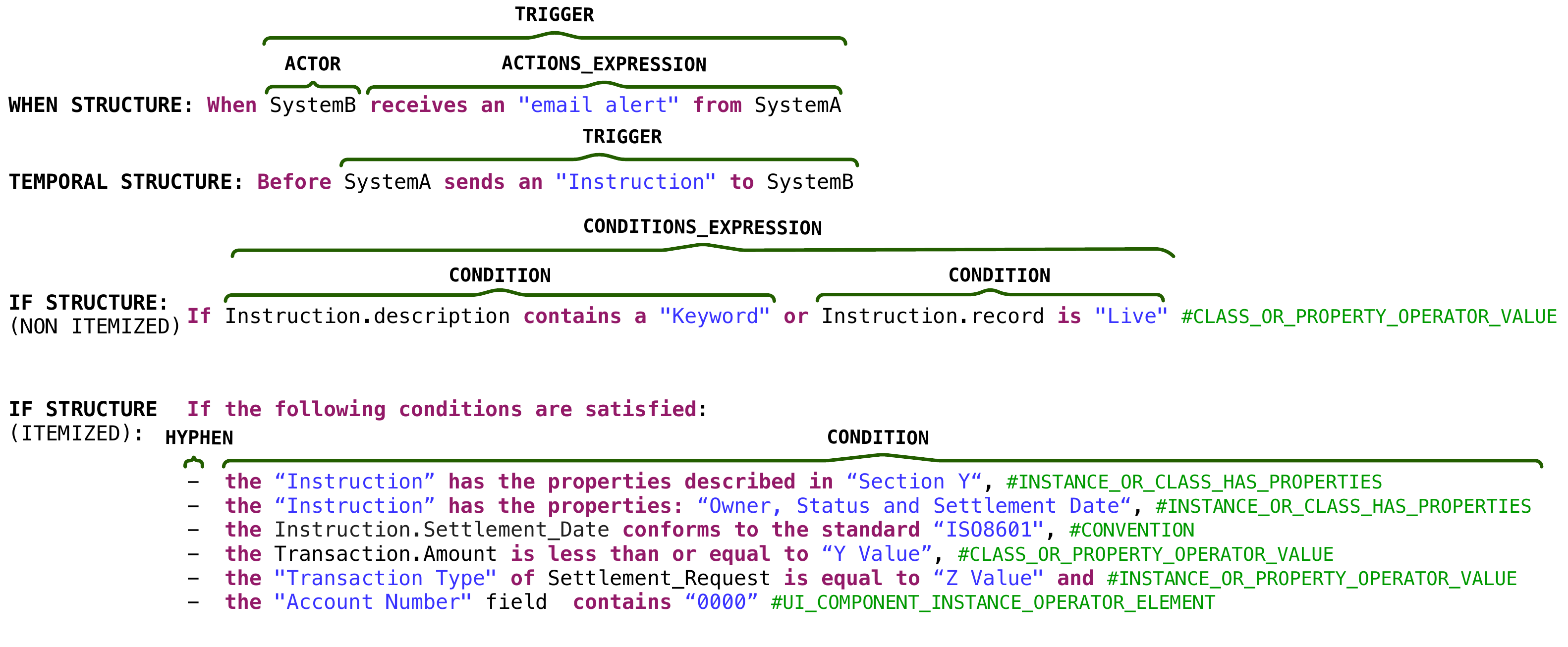}}
	\caption{Examples of condition structures}
	\label{fig:ExampleConditionStructure}
\end{figure}
\end{landscape}

\subsubsection{Condition Rule}
The operators and operands defined in the previous subsections are used in the five grammar rules shown in Listing~\ref{GeneralGrammar_C} conditions such as the ones shown in Figure~\ref{fig:ExampleConditionStructure}. 

\begin{lstlisting}[frame=single,
label={GeneralGrammar_C},
linerange=3-16, 
language=OwnExamples,
caption={Conditions rules}]
INSTANCE_OR_CLASS_HAS_PROPERTIES: MODIFIER? (INSTANCE|CLASS) CONTAINS (the properties described in TEXT)|(the (property|properties): PROPERTIES)

CONVENTION: ((TEXT (of CLASS)? | PROPERTY) (conforms|conform|comply|complies) (to|with) MODIFIER? (format|convention|standard) TEXT 

CLASS_OR_PROPERTY_OPERATOR_VALUE: MODIFIER? CLASS|PROPERTY OPERATOR_VALUES_EXPR

INSTANCE_OR_PROPERTY_OPERATOR_VALUE:	MODIFIER? TEXT OF_CLASS_OR_REFERENCE_TO_LABEL? Label? OPERATOR_VALUES_EXPR

UI_COMPONENT_INSTANCE_OPERATOR_ELEMENT: MODIFIER? TEXT UI_COMPONENT OPERATOR_VALUES_EXPR

OPERATOR_VALUES_EXPR: (COMPARE|CONTAINS|OTHER_OPERATORS) MODIFIER? MULTI_VALUES_EXPR TEXT?


\end{lstlisting}

The types of conditions are described as follows:
 
 (1) \texttt{INSTANCE OR CLASS HAS PROPERTIES} evaluates if the instance of a class, or a class itself defines one or more specific properties. The properties can be defined in a document (e.g., ``\CNL{Instruction has the properties described in the Section 1.b}"), or directly in the requirement (e.g., ``\CNL{Instruction has the properties: Owner, Status and Settlement_Date}").

(2) \texttt{CONVENTION} checks if a property conforms to a format or standard, e.g., ``\CNL{Instruction.Settlement_Date conforms to the standard ISO-8601}".

(3) \texttt{CLASS OR PROPERTY OPERATOR ELEMENT} is a condition composed of an operand-1, an operator and an  operand-2. The operand-1 is a reference to a \texttt{CLASS} or \texttt{PROPERTY}. The auxiliary rule \texttt{OPERATOR VALUES EXPR} defines the operator and the operand-2 of the condition, e.g., ``\CNL{the Transaction.Amount is less than or equal to 20000 Euros}''. The operand-2 is any type of operand described in Section~\ref{subsubsec:Operands}.

(4) \texttt{INSTANCE OR PROPERTY OPERATOR VALUE} is an operand-operator-value condition. 
The operand is a reference to an  \texttt{INSTANCE} or  \texttt{PROPERTY} and the value represent any literal or number.
An example of this type of condition is: ``\CNL{Transaction Type of Settlement Request is equal to Z-Value}".

(5) \texttt{UI COMPONENT INSTANCE OPERATOR ELEMENT} is a condition composed by an operand-1, operator, and operand-2 for a requirement related to the user interface~(UI). The operand-1 is an instance of a UI component identified by a free form \texttt{TEXT} followed by a reference to the type of \texttt{UI COMPONENT}. \ApproachName contains a list of common UI component types to help the user to create the requirements (e.g., tab, page, bar, field, calendar, checkbox, menu, message). The auxiliary rule \texttt{OPERATOR VALUES EXPR} defines the operator and the operand-2 of the condition. An example that displays this type of condition is: ``\CNL{the Account Number field contains 0000}".

\subsection{System Response}\label{subsec:systemResponse}

The rule \texttt{SYSTEM\_RESPONSE} in Listing~\ref{GeneralGrammar_D} allows the user to express the behavior of the system in two manners using the rules: (a) \texttt{RESPONSE\_BLOCK\_ITEMIZED}, that  is suitable for writing lists of actions; and (b) \texttt{SYSTEM\_RESPONSE\_EXPRESSION}, that is appropriate for writing one or multiple actions combined with logical operators, or parentheses that denote the priority of the actions. The previous rules include the rule \texttt{ATOMIC\_SYSTEM\_RESPONSES} and logical operators. Each \texttt{ATOMIC\_SYSTEM\_RESPONSE} contains an \texttt{ACTION\_PHRASE} and optionally, a frequency (e.g.,  \CNL{every 3 seconds}). 
\begin{lstlisting}[frame=single,
label={GeneralGrammar_D},
linerange=1-6, 
language=OwnExamples,
caption={System response}]
SYSTEM_RESPONSE: SYSTEM_RESPONSE_EXPRESSION | RESPONSE_BLOCK_ITEMIZED
ATOMIC_SYSTEM_RESPONSE: ACTION_PHRASE (every TEXT )?
RESPONSE_BLOCK_ITEMIZED: do the following actions (in sequence)? : 
	BULLET ATOMIC_SYSTEM_RESPONSE ((,|, and|, or|and|or)? 
	BULLET ATOMIC_SYSTEM_RESPONSE) *
		
\end{lstlisting}

All the types of \texttt{ACTION\_PHRASE} rules are available in \mbox{Appendix~\ref{sec:Grammar-Appendix}}. The rule \texttt{OBTAIN\_13\_5\_2} in Table~\ref{tab:GrammarRuleObtain} is one type of \texttt{ACTION\_PHRASE} rule. The column ``Grammar Rule Name" shows the name of the grammar rule related to the code \emph{obtain 13.5.2} that we discovered during the qualitative study (Tables~\ref{table:CodingResultsVNetQualitative} and \ref{table:CodingResultsProposedQualitative}). The column ``Grammar Rule Summary" describes the syntax of \texttt{OBTAIN\_13\_5\_2}, and the column ``Examples" shows requirements that conform to that syntax.

\begin{table}[htpb]
\centering
\caption{Grammar rule: \texttt{OBTAIN\_13\_5\_2}}
\label{tab:GrammarRuleObtain}
\begin{tabular}{p{1.7cm} | p{4.5 cm} | p{4.5cm} } 
\hline
\textbf{Grammar Rule Name} & \hfil \textbf{Grammar Rule Summary} & \hfil \textbf{Examples} \\

\hline
\texttt{OBTAIN\_13\_5\_2}  & 
\tinyCNL$accept|receive|retrieve|reject $
\tinyCNL$MODIFIER? INSTANCE | CLASS $
\tinyCNL$(from ELEMENTS)? $
\tinyCNL$(through ACTORS)? $
\tinyCNL$(in compliance with TEXT $
\tinyCNL$(described in TEXT)?)?$
& \exCNL$Example 1: receive a DA_file from CFCL_IT$
\newline \exCNL$Example 2: reject the "Message" in compliance with "current validation rules"$\\
\hline
\end{tabular}
\end{table}

\subsubsection{\ApproachName Editor}
\begin{figure}[htpb]
\centerline{\includegraphics[width=\textwidth]{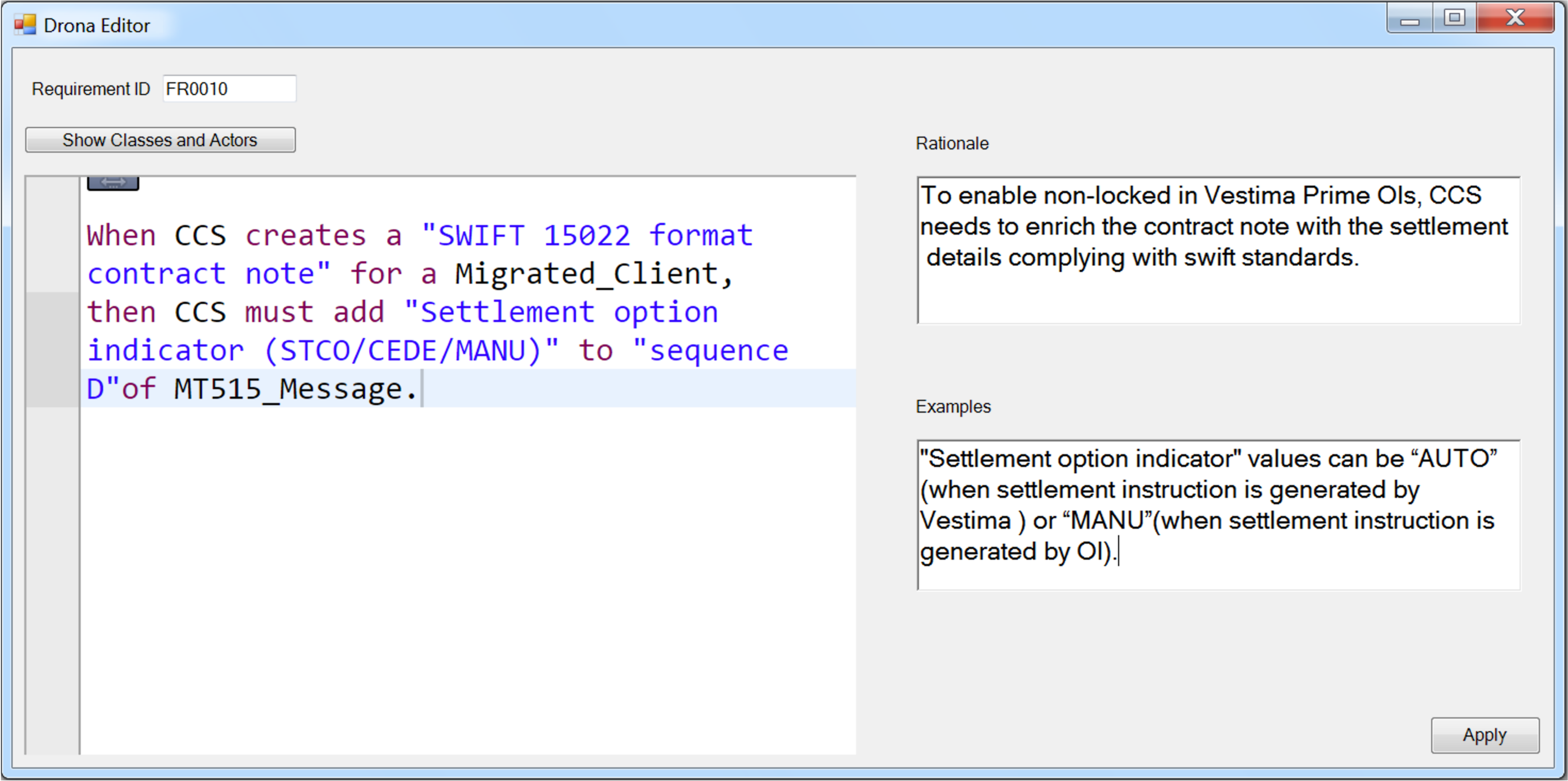}}
\caption{Screenshot of the requirements entry dialog box in the \ApproachName editor}
\label{fig:tool}
\end{figure}

We developed the \ApproachName editor using the Xtext language engineering framework~\citep{bettini2016implementing} which enables the development of textual domain-specific languages. We integrated the \ApproachName editor into an existing and widely known modeling and code-generation tool: Sparx Systems Enterprise Architect\footnote{\url{https://sparxsystems.com/products/ea/}}. Enterprise Architect was already being used at Clearstream. In particular, we created a form composed of the \ApproachName editor, and fields related to key properties of a requirement, such as 
``Requirement ID",
``Rationale", and ``Examples". Figure~\ref{fig:tool} shows a screenshot of the form.

To operationalize our technology-independent grammar (created in Step~5), we need to enhance it with some additional information. In particular, Xtext requires one to declare the name of the language, and further, import reusable terminals such as \emph{INT}, \emph{STRING} and \emph{ID} for the syntax of integers, text, and identifiers, respectively.

The input that we provided to Xtext is an EBNF-like grammar composed of rules that are similar to the ones that we discussed in this section. Xtext automatically generates a web-based editor with the following helpful features~\citep{bettini2016implementing}: (a) syntax highlighting, it allows to have the requirements colored and formatted with different visual styles according to the elements of the language; (b) error markers, when the tool automatically highlights the parts of the requirements indicating errors; and (c) content assist, a feature that automatically, or on demand, provides suggestions to the financial analysts on how to complete the statement/expression. In practice, these features are important to facilitate the adoption of \ApproachName by financial analysts.

\noindent\fbox{%
    \parbox{\textwidth}{
\textbf{Answer to RQ2:} We operationalized the grammar of \ApproachName developed in Section~\ref{sec:QualitativeStudy} into a full-featured editor using Xtext. Nevertheless, \ApproachName is independent of any language engineering framework. Our grammar offers broad coverage of system response and condition types, following recommended syntactic structures for requirements (e.g., the use of active voice). 
    }%
}
\section{Empirical Evaluation}\label{sec:EmpiricalEvaluation}
In this section, we describe a case study that evaluates \ApproachName developed in Sections~\ref{sec:QualitativeStudy} and \ref{sec:Grammar}. 
Throughout the section, we follow best practices for reporting on Case Study Research in Software Engineering~\citep{Runeson:2012}.

\subsection{Case Study Design}
\label{subsec:CaseStudy}
\begin{figure}[htpb]
\centering
\centerline{\includegraphics[width=\linewidth]{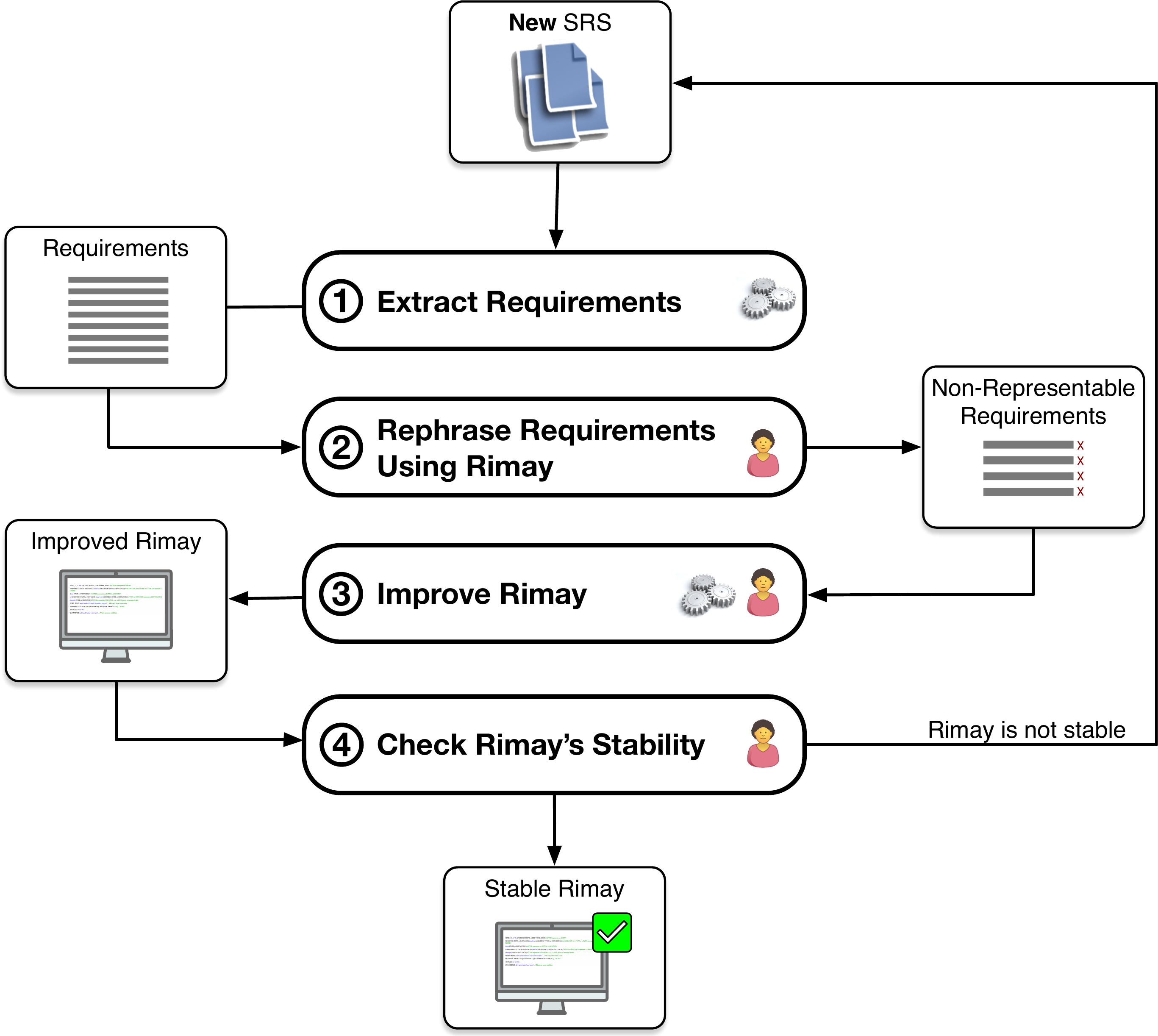}}
\caption{Case study design}
\label{fig:EvaluationProcess}
\end{figure}

As stated in the introduction, our evaluation aims to answer the following research questions:
\begin{itemize}

\item \textbf{RQ3: How well can \ApproachName express the requirements of previously unseen documents?}
\item \textbf{RQ4: How quickly does \ApproachName converge towards a stable state?}
\end{itemize}
Figure~\ref{fig:EvaluationProcess} shows the iterative process that we follow in order to answer these two questions. To evaluate our approach, we needed to collect new SRSs that had not been used for the construction of \ApproachName. 
We applied the four steps presented in Figure~\ref{fig:EvaluationProcess} to collect new SRSs and examine the expressiveness and stability of \ApproachName using them: \textbf{(Step~1)}~The financial analysts, on an opportunistic basis, gave us a new SRS that we had not seen before; we extracted from the given SRS its NL requirements (``Extract Requirements", Section~\ref{subsec:ExtractRequirements}). \textbf{(Step~2)}~We attempted to rephrase the extracted requirements using the rules of \ApproachName, keeping the intent of the original requirements and ensuring that we did not lose any information content. In this step, we had to keep track of the requirements, if any, that were \emph{non-representable} as well as the causes for such limitations (``Rephrase Requirements Using \ApproachNameNoSpace", Section~\ref{subsec:RephraseRequirements}). \textbf{(Step~3)}~We analyzed the requirements that were marked as \emph{non-representable} and enhanced \ApproachName   
to make these requirements \emph{representable} 
(``Improve \ApproachNameNoSpace", Section~\ref{subsec:ImproveGenerate}). \textbf{(Step~4)}~We checked whether there was a significant change in \ApproachName's ability to capture previously unseen content. As we argue in Section ~\ref{subsect:saturation}, it turned out that with four SRSs (i.e., four iterations of the process in Figure~\ref{fig:EvaluationProcess}), we were able to reach saturation. At that point, we stopped analyzing more SRSs (``Check \ApproachName's Stability", Section~\ref{subsec:CheckStability}). In the remainder of this section, we will not repeatedly be stating that these four SRSs were collected and analyzed iteratively and in a sequence. Instead, for succinctness, we refer to these four SRSs collectively when it is more convenient to do so.

With regard to our research questions, Step~1 and Step~2 of the process in Figure~\ref{fig:EvaluationProcess}  answer RQ3, as these two steps provide information about the expressiveness of \ApproachName, i.e., the requirements that were \emph{representable} or \emph{non-representable} with \ApproachName. Step~3 and Step~4 of the process address RQ4, as these steps provide information about the improvements necessary for maturing \ApproachName to a stable state.

\subsubsection {Extract Requirements (Step~1 of Figure~\ref{fig:EvaluationProcess})}
\label{subsec:ExtractRequirements} 

In Step~1 of Figure~\ref{fig:EvaluationProcess}, we extract the requirements from our four new, previously unseen SRSs. These SRSs were selected by senior financial analysts from Clearstream according to the criteria described in  Section~\ref{subsec:StudyContexDataSelection}. The selected SRSs did not contain any requirement that was already analyzed while building \ApproachName's grammar in the qualitative study of Section~\ref{sec:QualitativeStudy}.

\subsubsection{Rephrase Requirements Using \ApproachName (Step~2 of Figure~\ref{fig:EvaluationProcess})}
\label{subsec:RephraseRequirements}

A team composed of two annotators (the first and second authors of this article) rephrased the requirements using \ApproachName. 
A requirement can be composed of a scope, pre-conditions, an actor, and a system response. The scope and pre-conditions are optional, but the presence of at least one system response and one actor is mandatory. 

Step~2 considers  a requirement to be \emph{non-representable} when some information content of the requirement cannot be captured using \ApproachName. 
A requirement is considered \emph{representable}, otherwise. A requirement that is \emph{non-representable} is annotated with one of following three causes:

\begin{itemize}
\item \emph{Cause 1}. The requirement contains a verb that is not supported by \ApproachName rules. Therefore, we can either extend a \ApproachName rule with the verb or create a new rule. 
\item \emph{Cause 2}. Part of the requirement (excluding the verb) includes information content that is not supported by \ApproachName. 
\item \emph{Cause 3}. The meaning of the requirement is unclear and no financial analyst could clarify it.
\end{itemize}

\subsubsection{Improve \ApproachName (Step~3 of Figure~\ref{fig:EvaluationProcess})}
\label{subsec:ImproveGenerate}
To improve \ApproachName, we analyzed the causes for requirements marked as \emph{non-representable}. Concretely, we enhanced \ApproachName grammar by: (a) creating a new grammar rule when such requirement was marked with $Cause~1$. To create a new grammar rule, we first identified, for each requirement, the codes according to the steps described in  Section~\ref{subsub:identify}. The resulting codes were either identified from VerbNet or proposed by us. We then created the grammar rules following the steps described in  Section~\ref{subsec:createGrammar}; and (b) updating an existing grammar rule created in Section \ref{sec:QualitativeStudy} to include either a new verb of a requirement labeled with $Cause~1$ or missing content of a requirement labeled with $Cause~2$.  

Requirements labeled with $Cause~3$ were not addressed in \ApproachName. We discuss such requirements in Section \ref{sec:ThreatsToValidity}, dedicated to threats to validity.

\subsubsection{Check \ApproachName's Stability (Step~4 of Figure~\ref{fig:EvaluationProcess})}
\label{subsec:CheckStability}

This step verifies whether there was a significant change in \ApproachName's capacity to capture the content of previously unseen NL requirements. If there is no significant change, we say that \ApproachName is stable, and we stop the evaluation process. Otherwise, we iterate over Step~1 to Step~4 using a new SRS until \ApproachName becomes stable.
We refer to the notion of \emph{saturation} to determine the point where \ApproachName is stable. 
We reach the saturation point when \ApproachName is expressive enough to capture all the verbs in the NL requirements of a SRS (i.e., the number of errors due to Cause~1 is zero).
In our case study, we reached the saturation point during the evaluation of SRS~4.


\subsection{Data Collection}
\label{subsec:DataCollection}
\begin{figure}[htpb]
\caption{Data model of the collected requirements}
\label{fig:DataModel}
\centering
\centerline{\includegraphics[width=\linewidth]{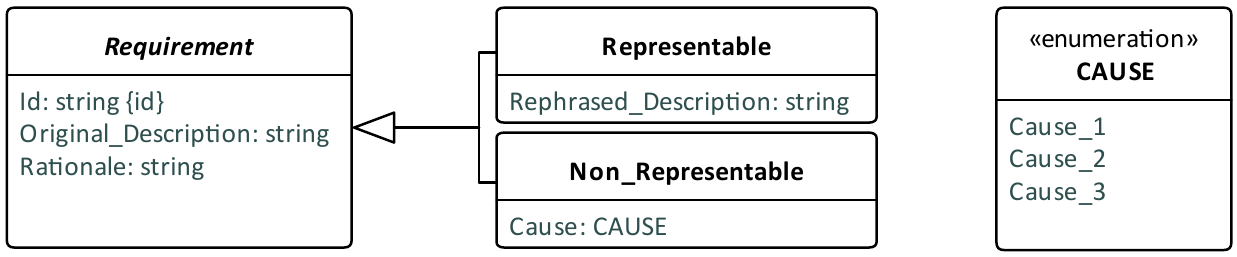}}
\end{figure}

We answered RQ3 and RQ4 by collecting data from the execution of the four steps described in Section~\ref{subsec:CaseStudy}. 
Figure~\ref{fig:DataModel} shows the data model of the requirements collected during the empirical evaluation. In our data model, a \var{Requirement} has an \var{Id} which is a unique code assigned to each requirement, an \var{Original\_Description} and a \var{Rationale}. A requirement is either \var{Representable} or \var{Non\_Representable}. If the requirement is \var{Representable}, we recorded its \var{Rephrased\_Description}. If the requirement is \var{Non\_Representable}, we recorded the \var{CAUSE} (i.e., \var{Cause\_1}, \var{Cause\_2} or \var{Cause\_3}).

In total, we collected 460 requirements from the four SRSs used in our evaluation. We improved the grammar rules after rephrasing one SRS and assessed the improved grammar on the next.

\subsection{Collecting Evidence and Results}
\label{subsec:Results}
This section describes the execution and the raw data collected from our case study. The case study required the work of two annotators for two months, adding up to approximately 200 person-hours.


Table~\ref{table:requirementsExpressibleUsingNLR} provides the data for each one of the four SRSs. For each SRS, we measure the percentage of requirements that can be represented using \ApproachName. For example, the first row of Table~\ref{table:requirementsExpressibleUsingNLR} shows that 74,7\% of the requirements (65 out of 87) of the first SRS are \emph{representable} with \ApproachName.

\begin{table}[htpb]
\centering
\caption{Percentage of \emph{representable} requirements and frequencies of causes for \emph{non-representable} requirements}
\label{table:requirementsExpressibleUsingNLR}
\begin{tabular}{ c | c | c | c | c |c }
\hline
&  &  \hfil \textbf{\%}& \multicolumn{3}{c}{\textbf{Frequencies of Causes in}}\\  
\textbf{SRS} &\textbf{Number of } & \emph{\textbf{Representable}}&\multicolumn{3}{c}{\textbf{\emph{Non-Representable} Requirements} }\\\cline{4-6} 
\textbf{ID}&\textbf{Requirements} &\textbf{Requirements}   &  \emph{\textbf{Cause 1}} & \emph{\textbf{Cause 2}} & \emph{\textbf{Cause 3}} \\

\hline
1  & 87      & \hfil 74,7    & 11 & 9 & 2 \\    
2  & 113     & \hfil 85,0    & 6 & 8 & 3 \\    
3 & 192      & \hfil 93,8    & 2 & 7 & 3 \\   
4 & 68       &\hfil  94,1    & - & 4 &-\\     
\hline
\end{tabular} 
\end{table}

Table~\ref{table:requirementsExpressibleUsingNLR} shows, for the four SRSs
, the frequency of the three causes (described in Section~\ref{subsec:RephraseRequirements}) in the requirements labeled as \emph{non-representable}. For example, the first row of Table~\ref{table:requirementsExpressibleUsingNLR}, i.e., SRS~1, shows that for 11 requirements, the verb was not supported by \ApproachName (\emph{Cause~1}). For nine requirements, some other content was not supported by \ApproachName (\emph{Cause~2}). Two requirements were unclear and no financial analyst could clarify them (\emph{Cause~3}). In total, 22 out of 87 requirements (25,3\%) in SRS 1 were \emph{non-representable.} 

Finally, we improved \ApproachName by addressing the \emph{non-representable} requirements labeled with Causes 1 and 2, as explained in Section  \ref{subsec:ImproveGenerate}. 

\subsubsection{Coding Results.}\label{subsec:codingResultsEvaluation}

Tables~\ref{table:CodingResultsVNetEvaluation} and \ref{table:CodingResultsProposedEvaluation} show the codes and their verb members identified during our empirical evaluation. Recall from Section~\ref{sec:QualitativeStudy} that a code represents a group of verbs that convey the same information in NL requirements. The structures of Table~\ref{table:CodingResultsVNetEvaluation} and Table~\ref{table:CodingResultsProposedEvaluation} are the same as the structures of Table~\ref{table:CodingResultsVNetQualitative} and Table~\ref{table:CodingResultsProposedQualitative} reporting the coding results of our qualitative study discussed in Section~\ref{sec:QualitativeStudy}.

Seven out of 13 codes in Tables~\ref{table:CodingResultsVNetQualitative} and~\ref{table:CodingResultsProposedQualitative} were found during our empirical evaluation. We placed the symbol ``*" before the seven new codes to differentiate them from the codes that we had already identified in the qualitative study. For each new code, we created a new grammar rule.
Considering that, in total, we found 48 codes during the qualitative study and the empirical evaluation, the seven (14,6\%) new codes found in the empirical evaluation did not prompt drastic modifications to \ApproachName.

\begin{table}[ht]
\centering
\caption{VerbNet codes identified during our empirical evaluation}
\label{table:CodingResultsVNetEvaluation}
\begin{tabular}{p{1.5cm} | l | p{1.9cm} }
\hline
\multicolumn{2}{c|}{\textbf{Codes}} 	 & \multicolumn{1}{c}{\multirow{2}{*}{ \textbf{Members}}}  \\[0.1cm] \cline{1-2}
\multicolumn{1}{c|}{ \textbf{Class name}} & \multicolumn{1}{c|}{\textbf{Hierarchy Level} } & \multicolumn{1}{c}{}                                   \\ \hline
begin & 55.1-1&start\\
$*$establish & 55.5-1 & establish \\
other\_cos & 45.4&reverse\\
remove & 10.1 & deduct \\
$*$search & 35.2 & search \\
send & 11.1 & export\\
$*$stop & 55.4 & stop\\
use & 105 & use \\
\hline 
\multicolumn{3}{c}{Total: 8} \tabularnewline
\hline
\end{tabular}
\end{table}

\begin{table}[ht]
\begin{minipage}{\textwidth}
\renewcommand\footnoterule{}
\centering
\caption{Codes proposed during our empirical evaluation}
\label{table:CodingResultsProposedEvaluation}
\begin{tabular}{ p{2.5cm} |  p{3.0cm}    }
\hline 
\textbf{Codes} & \textbf{Members} \\  [0.1cm]
\hline
$*$calculate & calculate, recalculate \tabularnewline
$*$split 		&split\\
$*$subscribe	&subscribe\\
$*$upload	&upload\\
update &set\\
\hline
\multicolumn{2}{c}{Total: 5}\tabularnewline
\hline
\end{tabular}
\end{minipage}
\end{table}
 
\subsection{Analysis of Collected Data}
\label{subsec:Analysis}
In this section, we analyze the collected data and answer RQ3 and RQ4.

\subsubsection{Performance of \ApproachName on Previously Unseen SRSs (RQ3)}
Table~\ref{table:requirementsExpressibleUsingNLR} shows that, on average, 88\% of  requirements (405 out of 460) can be expressed using \ApproachName. 
With regard to SRS~1, we note that we found five occurrences of a new verb, ``use", which we had not encountered during our qualitative study. The relatively low expressiveness in this first SRS is largely explained by the high frequency of appearance of this single verb. 
As one can see from Table~\ref{table:requirementsExpressibleUsingNLR}, most requirements can be represented in \ApproachName across all  SRSs. 
The improvements to the expressiveness of \ApproachName are brought about by small changes to \ApproachName. In other words, while the expressiveness of our grammar did improve as the result of analyzing more SRSs, we did not have to make major changes to the grammar. Our changes involved only the introduction of a few new verbs (as shown in Tables~\ref{table:CodingResultsVNetEvaluation} and \ref{table:CodingResultsProposedEvaluation}), and the enhancements of a small number of grammar rules created during our qualitative study (Section \ref{sec:QualitativeStudy}).

The most common causes for a requirement to be \emph{non-representable}, in order of prevalence, are \emph{Cause~2} with 28 occurrences (50.9\%), followed by \emph{Cause~1} with 19 occurrences~(34.5\%), and, finally \emph{Cause~3} with 8 occurrences~(14.5\%). We conjecture that the main reason why \emph{Cause~2} turns out to be the most frequent cause is that VerbNet -- the lexicon we use for deriving our grammar rules -- is domain-independent and may not contain certain information content that is specific to the financial domain. During our qualitative study, we identified some new content and extended the grammar rules accordingly. 
For example, the syntax for the rule \texttt{Send\_11.1} in VerbNet specifies that an \texttt{AGENT} can move a \texttt{THEME} (e.g., data) from an \texttt{INITIAL\_LOCATION} to a \texttt{DESTINATION}. Then, during the qualitative study, we identified new information content such as the temporal structure (e.g., ``Before 1h00 CET") used at the beginning of requirements. Furthermore, in the evaluation, we identified extra information content such as a valid channel to send the \texttt{THEME} (e.g., a subsystem that encrypts the data).
\\[12pt]
\noindent\fbox{
    \parbox{\textwidth}{
\textbf{Answer to RQ3: }\ApproachName performed well in expressing the requirements of unseen SRSs. In particular, 405 out of the 460 requirements (i.e., 88\%) used in our empirical evaluation were successfully rephrased using \ApproachName. The expressiveness of \ApproachName did steadily improve and converged to 94\% in the last SRS. The rephrased requirements maintained their original intent with no information loss.   We observed that improving the expressiveness of \ApproachName involved only small changes to its grammar.  This suggests that the version of \ApproachName obtained from our qualitative study (Section \ref{sec:QualitativeStudy}) did not require drastic changes to maximize expressiveness.  
    }
}

\subsubsection{Ensuring the Stability of \ApproachName (RQ4)}\label{subsect:saturation}
We refer to the notion of \emph{saturation} to determine the point in our evaluation where we have been through enough SRSs to be confident that the updated version of \ApproachName is as expressive as possible to specify requirements for the financial domain.  To determine if a statistically significant change is observed in the percentage of \emph{representable} requirements, we conduct z-tests for differences in proportions of representability across different SRSs. 


\paragraph{Saturation.} Usually, saturation is reached in a qualitative study when ``no new information seems to emerge during coding, i.e., when no new properties, dimensions, conditions, actions/interactions, or consequences are seen in the data"~\citep{GlaStr67}. In our evaluation, the saturation point is reached when all the verbs analyzed in a SRS are already considered by \ApproachName (i.e., when \emph{Cause~1} is not triggered). Specifically, 
as shown in Table~\ref{table:requirementsExpressibleUsingNLR}, SRS 4 was the only SRS where no requirement was classified as \emph{non-representable} due to \emph{Cause~1}. 

As can be seen from Table~\ref{table:requirementsExpressibleUsingNLR}, the increment in the percentage of requirements that can be written in \ApproachName is tangible  evidence that the changes made to \ApproachName were beneficial (although not extensive). 


\paragraph{Z-test.}
The z-test is a standard statistical test used for checking the difference between two proportions~\citep{Dietterich98}. We run one-tailed z-tests to check if the proportion ($p_1$) of \emph{representable} requirements in one SRS (SRS $i$) is larger than or equal to the proportion ($p_2$) of \emph{representable} requirements in another SRS (SRS $j$) analyzed thereafter. 
Our null and alternative hypotheses are as follows:
\[H_0 : p_1 \geq p_2 \]
\[H_1 : p_1 < p_2 \]
\begin{itemize}
    \item \texttt{$H_0:$} The percentage of \emph{representable} requirements does not increase from SRS~$i$ to SRS $j$.
    \item \texttt{$H_1:$} The percentage of \emph{representable} requirements increases from SRS~$i$ to SRS~$j$.
\end{itemize}
In total, we run six z-tests, at a level of significance of 0.05. The SRS pairs covered by these tests, alongside the corresponding proportions, are shown in Table~\ref{table:Z-Test_Inputs}. 
For example, the first row of Table~\ref{table:Z-Test_Inputs} shows the input for performing a z-test over the (SRS~1, SRS~2) pair. SRS~1 contains 65 requirements that are \emph{representable} with \ApproachName out of 87 requirements, and SRS~2 contains 96 requirements that are \emph{representable} with \ApproachName out of 113 requirements.
\begin{table}[ht]
\centering

\begin{tabular}{p{0.6cm}|c | p{1.1cm} | p{1.1cm} |p{2.3cm}|p{2.3cm}}
\hline
\textbf{Test} & \multicolumn{5}{c}{\textbf{Input}}\\\cline{2-6} 
 & \textbf{Document Pair} & \textbf{Sample} & \textbf{Sample}& \textbf{\emph{Representable}}&\textbf{\emph{Representable}} \\
 &\textbf{SRS $i$, SRS $j$}&\textbf{Size in}&\textbf{Size in}&\textbf{Requirements}&\textbf{Requirements}\\
 &&\textbf{SRS $i$}&\textbf{SRS $j$}&\textbf{in SRS~$i$}&\textbf{in SRS~$j$}\\
 &&&&\hfil \textbf{(\boldmath{$p_1$})}&\hfil \textbf{(\boldmath{$p_2$})}\\
\hline 

\hfil 1 & SRS 1, SRS 2 & \hfil 87 & \hfil 113 & \hfil 65 & \hfil 96\tabularnewline
\hfil 2 & SRS 1, SRS 3 & \hfil 87 & \hfil 192 & \hfil 65 & \hfil 180\tabularnewline
\hfil 3 & SRS 1, SRS 4 & \hfil 87 & \hfil 68 & \hfil 65 & \hfil 64\tabularnewline
\hfil 4 & SRS 2, SRS 3 & \hfil 113 & \hfil 192 & \hfil 96 & \hfil 180\tabularnewline
\hfil 5 & SRS 2, SRS 4 & \hfil 113 & \hfil 68 & \hfil 96 & \hfil 64\tabularnewline
\hfil 6 & SRS 3, SRS 4 & \hfil 192 & \hfil 68 & \hfil 180 & \hfil 64\tabularnewline
\hline
\end{tabular}
\caption{Z-tests inputs}
\label{table:Z-Test_Inputs}
\end{table}

The z-scores and p-values for the z-tests are shown in Table~\ref{table:Z-Tests Results}. We conclude that the null hypothesis, $H_0$, is \emph{rejected in the first five z-tests}. Therefore, there is significant evidence to claim that proportion $p_1$ is less than proportion $p_2$ at the 0.05 significance level for the first five document pairs. Concretely, this means that the proportion of \emph{representable} requirements in SRS~2, SRS~3, and SRS~4 are significantly better than that of SRS~1. Similarly, the proportion of \emph{representable} requirements in SRS~3 and SRS~4 are significantly better than that of SRS~2. However,  the null hypothesis \emph{cannot be rejected in the last z-test}. Therefore,  the proportion of \emph{representable} requirements in SRS~4 is not significantly better than that of SRS~3. We therefore concluded our analysis of new SRSs after completing SRS~4.

\begin{table}[ht]
\centering

\begin{tabular}{ c|c|c|c }
\hline 
\textbf{Test}& \textbf{Document Pair} & \textbf{z} & \boldmath{ $p-value$ }\\
& \textbf{SRS $i$, SRS $j$} & &\\

\hline 
1 &SRS 1, SRS 2& \textminus 1,81 & 0,03\tabularnewline
2 &SRS 1, SRS 3& \textminus 4,50 & 3,35 E-06\tabularnewline
3 &SRS 1, SRS 4 & \textminus 3,21 & 6,67 E-4\tabularnewline 
4 &SRS 2, SRS 3 & \textminus 2,53 & 0,01\tabularnewline 
5 &SRS 2, SRS 4 & \textminus 1,86 & 0,03\tabularnewline 
6 &SRS 3, SRS 4 & \textminus 0,11 & 0,46\tabularnewline 
\hline 
\end{tabular}

\caption{Z-test results}
\label{table:Z-Tests Results}
\end{table}

\noindent\fbox{%
    \parbox{\textwidth}{%
\textbf{Answer to RQ4:} We reached a stable version of our grammar after analyzing SRS~3 in our evaluation set. During the analysis of SRS~4, no new verbs emerged; we therefore concluded that we had reached saturation. Statistical tests confirmed that, after analyzing SRS~3, changes to \ApproachName did not bring about significant improvements in expressiveness.
    }%
}
\section{Threats to Validity} \label{sec:ThreatsToValidity}
In the following subsections, we analyze potential threats to the  validity of our empirical work according to the categories suggested by~\citet{Wohlin:2012} and adapted by~\citet{Runeson:2012} for case studies in software engineering.

\subsection{Construct Validity}
Construct validity reflects to what extent the operational measures that are studied really represent what the researcher has in mind and what is investigated according to the research questions~\citep{Runeson:2012}.

We measured the percentages of the requirements that can be represented with \ApproachName according to the grammar rules we identified. If the criteria that we used to assess whether a requirement is \emph{representable} are incomplete or too strict, this could constitute a threat. We therefore proposed three criteria (named \emph{Causes}) that alleviate the risk of introducing inadequate information content into \ApproachName. 
We analyzed the \emph{Causes} of the requirements marked as \emph{non-representable} in order to enhance the \ApproachName grammar by (a) creating new grammar rules (i.e., \emph{Cause~1}); (b) updating grammar rules to include some missing content (i.e., \emph{Cause 1} and \emph{Cause~2}), and (c) not considering incomplete, ambiguous or unclear information content (i.e., \emph{Cause~3}). \emph{Cause~1} and \emph{Cause~2} are meant to capture missing parts that need to be included in the \ApproachName grammar. On the other hand, \emph{Cause~3} focuses on the requirements that describe incorrect information content that we do not want to include in \ApproachName. 
To be sure that no important information was excluded from \ApproachName, we looked at the eight \emph{non-representable} requirements labelled with \emph{Cause~3} (Table \ref{table:requirementsExpressibleUsingNLR}) with the senior financial analysts from Clearstream, who agreed with our decision to discard them.
\subsection{Internal Validity}
Internal validity is of concern when causal relations are examined~\citep{Runeson:2012}. 

The results and the conclusions of our study strongly rely on two key activities that were performed manually: (1) the identification of codes and their members, and (2) the transformation process of requirements into \ApproachName. This can represent an important threat to the internal validity of our study. To mitigate biases, these two activities were systematically performed by a pair of researchers (the first and second author of this article). Afterward, a third researcher (the third author of this article) reviewed and challenged some of the results of these activities. We finally improved steps (1) and (2) upon reaching an agreement between these three researchers.

Another threat to the internal validity is related to the assumption that all the requirements in SRSs should be used to create \ApproachName. If all the requirements in SRSs are used, incomplete and unclear requirements might be easily misinterpreted and as a consequence, incorrect information content might be included in \ApproachName. To tackle this threat, in Step~2 ``Rephrase Requirements Using \ApproachName" (Figure~\ref{fig:EvaluationProcess}), we first classified as \emph{non-representable due to Cause~3} the requirements that contained either incomplete or unclear information and we then discarded those requirements. 

\subsection{Reliability Validity}
Reliability validity is concerned with the extent to which the data and the analysis are dependent on the specific researchers involved~\citep{Runeson:2012}. In order to achieve acceptable reliability, research steps must be repeatable, i.e., other researchers have to be able to replicate our results~\citep{Badampudi2016}. 

It is impossible to build a CNL that is able to represent all software requirements, and as we already acknowledged, some requirements could not be represented with \ApproachName. The main issues that may constitute a threat to reliability are related to how we built our CNL to be as expressive as possible. To mitigate this threat, we described in details the steps of our qualitative study and empirical evaluation following a systematic process. This process was performed by the first and second authors and monitored by the other authors of the article.

\section{Practical Considerations}\label{sec:PracticalConsiderations}
In this section, we present some practical considerations for the different audiences who may be interested in the work reported in this article. These considerations are based on both our experience and our interactions with our industrial partner.

\textbf{Considerations for CNL builders.} 
The creation of a language editor entails a significant level of effort because there are many tasks to support, such as auto-completion and syntax highlighting. Mature language engineering frameworks make these tasks less complicated or even fully automated. For instance, we used Xtext to generate a basic editor based only on the grammar of \ApproachName. For us, the most challenging part of defining a grammar was to understand how to model nested expressions.
The effort to customize the generic behavior of the editor generated by Xtext should be considered. In our case, we use the generic editor for our evaluation, but we are in the process of customizing the editor to further improve usability. In particular, we are simplifying the error messages shown by \ApproachName's editor, since they are difficult to understand for people without technical knowledge.

\textbf{Considerations for companies investing into a CNL.}
Additional effort is to be anticipated for integrating a CNL with existing software development tools. 
In our case, our industrial partner uses Sparx Systems Enterprise Architect for modeling UML Use Case, Class, and Activity Diagrams. A key consideration for our partner was therefore to be able to reference (from requirements) the elements of UML models in Enterprise Architect. 
To provide such functionality, \ApproachName's editor dynamically tracks the model elements that need to be referenceable from requirements. This allows \ApproachName's editor to provide context-sensitive auto-completion assistance as analysts type in their requirements. Furthermore, if an analyst introduces in a requirement an element that does not already exist in the UML model, our editor will notify the analyst, asking whether the new element should be added to the UML model.

Whether an organization should invest into a CNL for requirements also depends on how requirements are elaborated and used within the organization. Generic text editing tools may suffice for analysts working on small projects. In our case, the types of projects our industrial partner is engaged in justified the construction of a CNL; the projects are not only large and complex but also involve multiple analysts from geographically dispersed locations. Systematic requirements writing practices that help mitigate incompleteness and ambiguity are thus key for our partner. In addition, the partner is interested in extracting accurate information from the requirements as a prerequisite step for automating such tasks as consistency checking between models and (textual) requirements, as well as generating test cases from requirements. Working toward such automation objectives would be very difficult without structured requirements, thus further justifying investment into a CNL.
\section{Conclusions}\label{sec:Conclusions}
In this article, we proposed a rigorous methodology to define controlled natural languages (CNLs) for requirements specifications. We applied this methodology to develop a CNL, which we named \ApproachName, for expressing functional requirements in the financial domain. 
\ApproachName's grammar was derived from a qualitative study based on the analysis of 2755 requirements from 11 distinct projects. In this qualitative study, we identified the information content that financial analysts should account for in the requirements of financial applications. 

%
We conducted an empirical evaluation of \ApproachName in a realistic setting. This evaluation measured the percentage of requirements that can be represented using \ApproachName. We observed that, on average, 88\% of the requirements that we evaluated in our case study (405 out of 460) could be expressed using \ApproachName. 
Additionally, we analyzed how quickly \ApproachName would converge and stabilize to even higher percentages when refined after each new requirements specification was analyzed.

The generalizability of our results is subject to certain limitations. For instance, we cannot conclude that \ApproachName is applicable to all the requirements in the financial domain, particularly because \ApproachName is based on the analysis of functional requirements only.

To a large extent, \ApproachName addresses the broader domain of data-intensive information systems.
We conjecture that many of our findings can be generalized to information systems in other similar domains, since, during our analysis, we did not resort to concepts that are specific to the financial domain, and instead, used domain-independent resources such as VerbNet and WordNet for constructing our grammar rules. 
That said, future investigations remain  necessary to determine whether and how \ApproachName can be specialized for other domains.

While CNLs and requirements patterns have generated a lot of attention in recent years as a vehicle for improving the quality of natural-language requirements, to our knowledge, no previous study has proposed and evaluated a CNL based on a qualitative analysis of a large number of industrial requirements and following a systematic process using lexical resources. A significant portion of this article was dedicated to developing and discussing such a systematic process with the goal of making this process repeatable; this way, other researchers and practitioners interested in developing their own CNLs can benefit from our proposed process and possibly even use \ApproachName as a starting point.



\paragraph{\textbf{Acknowledgements}} This project has received funding from Clearstream, Escent, FNR of Luxembourg under the BRIDGES program (grant BRIDGES18/IS/13234469/IMoReF), and NSERC of Canada under the Discovery, Discovery Accelerator and CRC programs.
\begin{appendices}
\section{Action Phrases in \ApproachName}\label{sec:Grammar-Appendix}
Table~\ref{tab:part1} and~Table~\ref{tab:grammarRulesEvaluation} show the name, summary, and examples of the \ApproachName grammar rules related to action phrases. Table~\ref{tab:part1}  displays the rules built during the qualitative study and Table~\ref{tab:grammarRulesEvaluation} depicts the rules created in the empirical evaluation.
\begin{table}[htp]
\centering
\includegraphics[width=\textwidth]{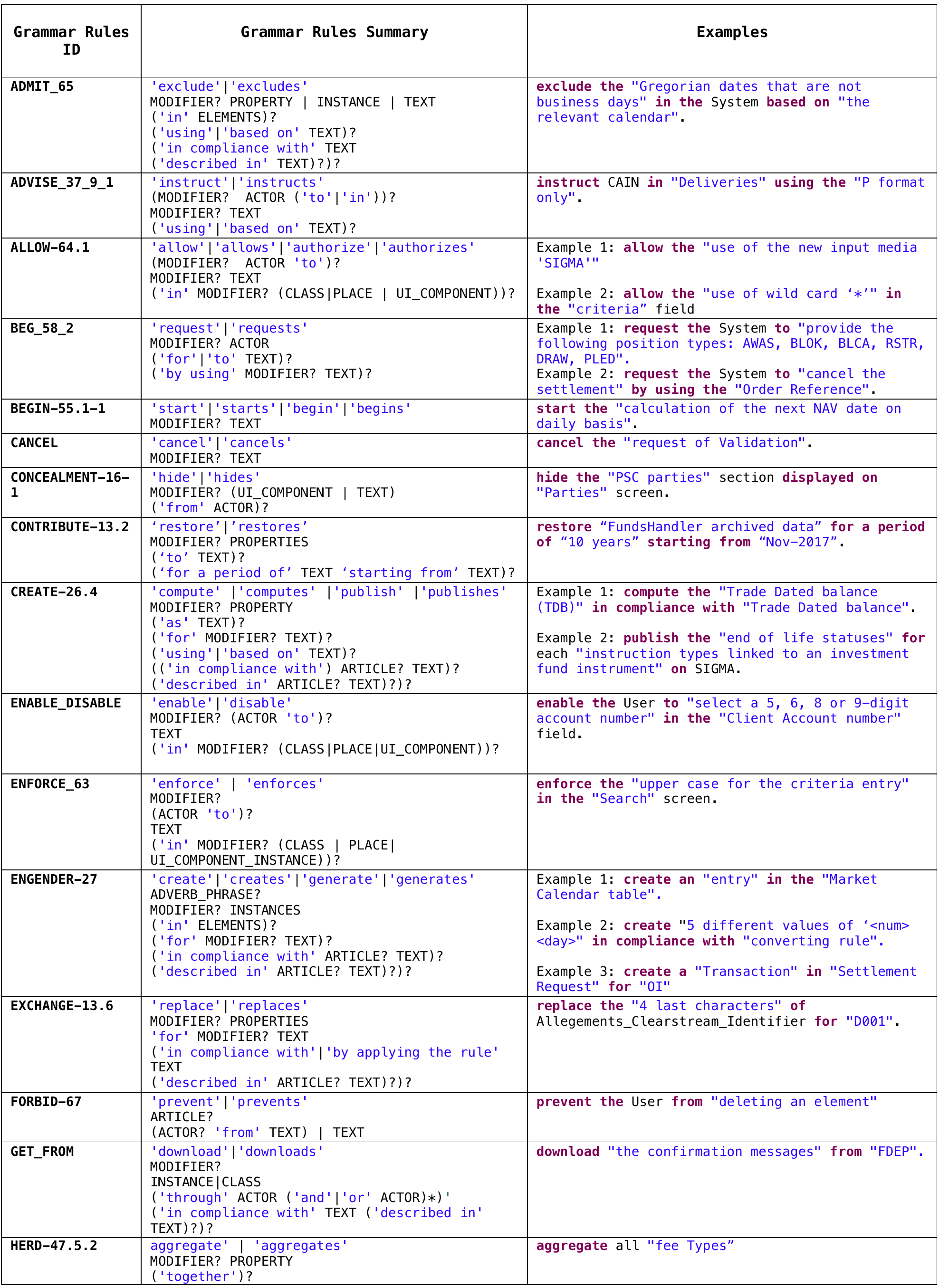}
\caption{Types of action phrase rules in \ApproachName (from Qualitative Study).}
\label{tab:part1}
\end{table}

\begin{table}[htp]
\centering
\ContinuedFloat
\includegraphics[width=\textwidth]{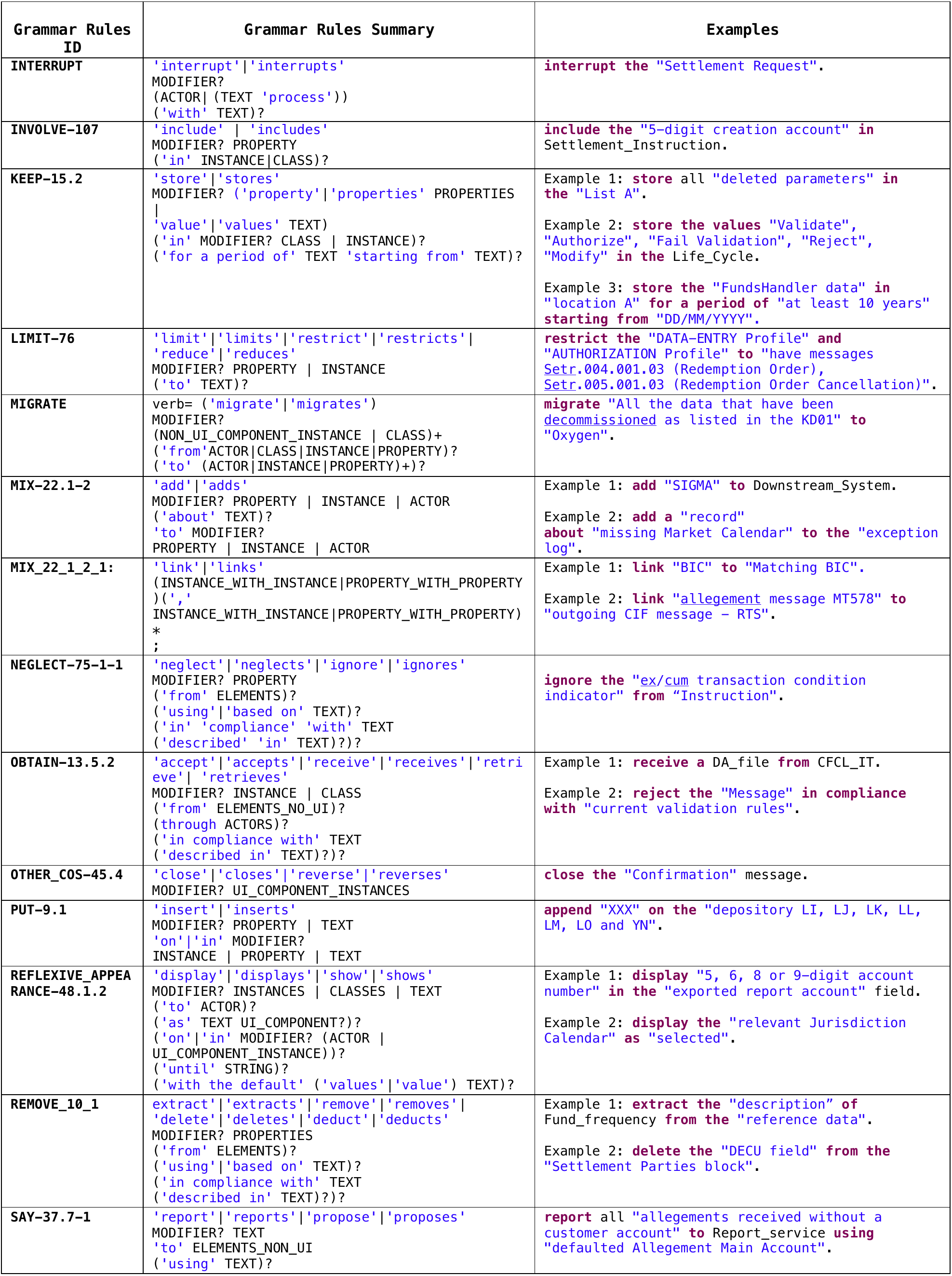}
\caption{(continued) Types of action phrase rules in \ApproachName (from Qualitative Study).}
\label{tab:part2}
\end{table}

\begin{table}[htp]
\centering
\ContinuedFloat
\includegraphics[width=\textwidth]{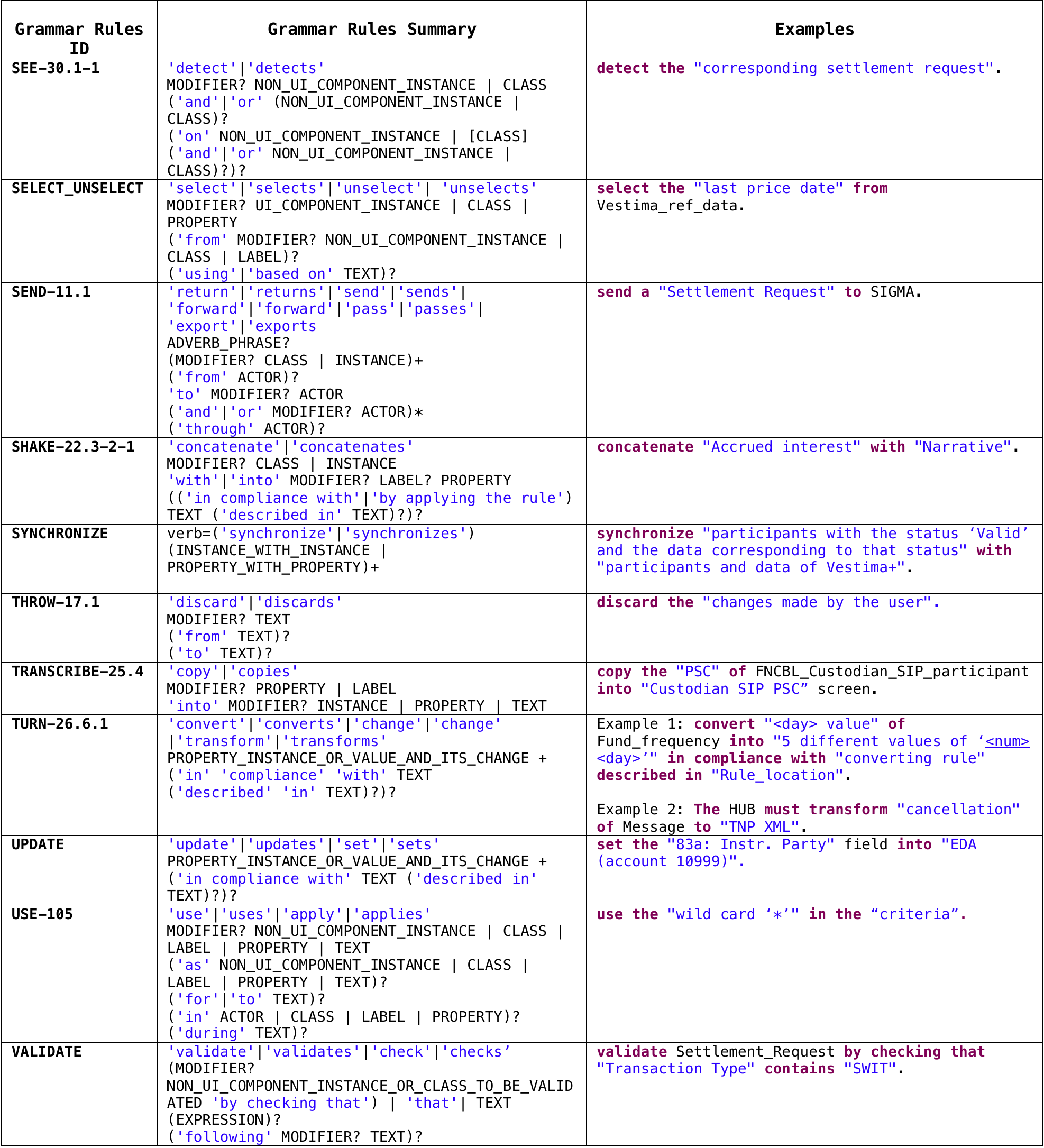}
\caption{(continued) Types of action phrase rules in \ApproachName (from Qualitative Study).}
\label{tab:part3}
\end{table}

\begin{table}[htp]
\centering
\includegraphics[width=\textwidth]{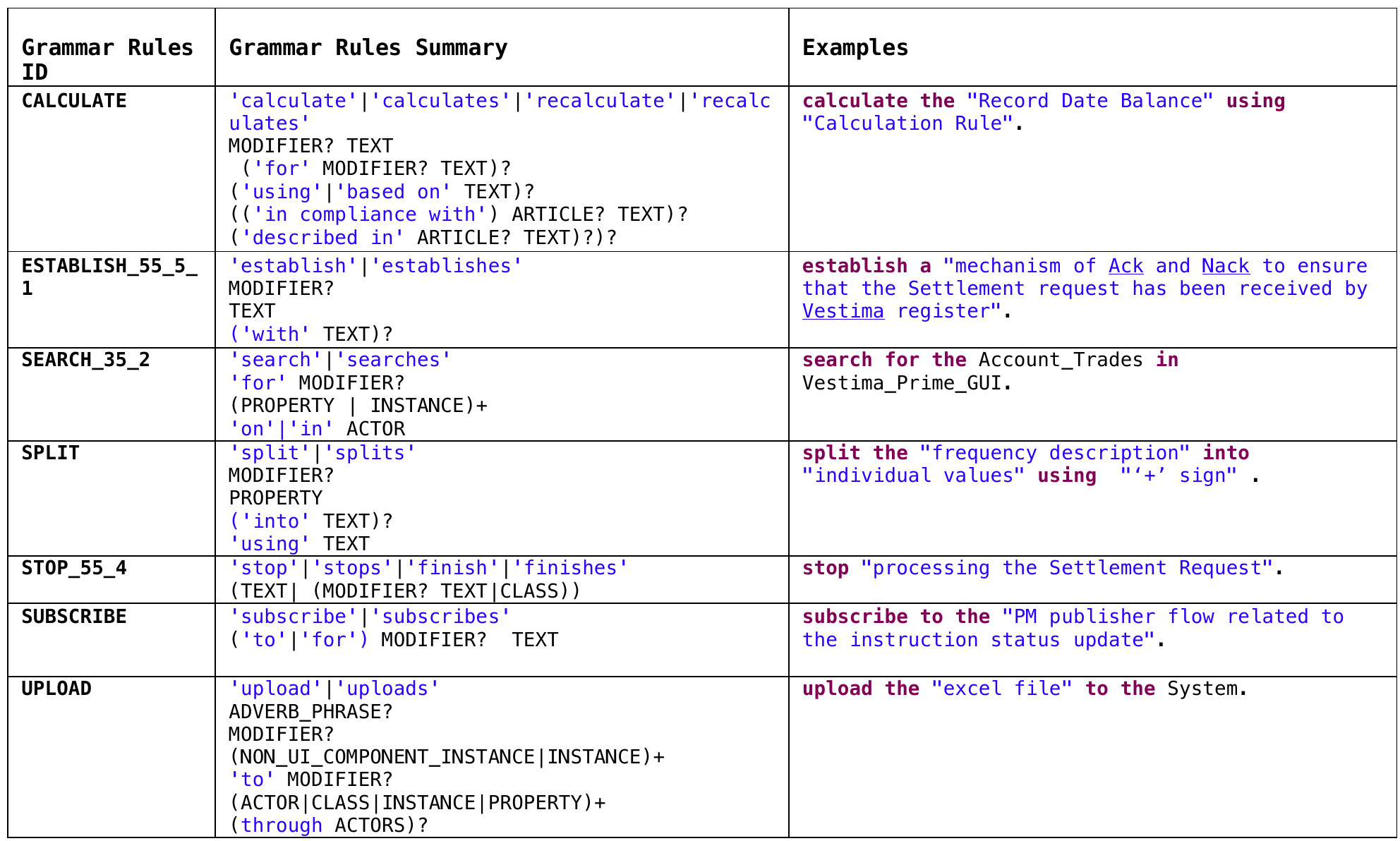}
\caption{Types of action phrase rules in \ApproachName (from Empirical Evaluation).}
\label{tab:grammarRulesEvaluation}
\end{table}

\label{table:CNLSection1}

\end{appendices}
\newpage


\bibliographystyle{spbasic}      
\clearpage
\bibliography{ref}   
\end{document}